\documentclass[10pt,journal]{IEEEtran}

\usepackage{graphicx}             
\usepackage{subfigure}            
\usepackage{url}                  
\usepackage[latin1]{inputenc}     
\usepackage{nicefrac}             
\usepackage{indentfirst}          
\usepackage{amsfonts}
\usepackage{amssymb}
\usepackage{algorithm}
\usepackage{algpseudocode}
\usepackage{multirow}

\newcommand{\dfn}[1]{\textit{#1}}            
\newcommand{\co}[2]{#1 & \tiny{$\pm#2$}}     

\graphicspath{{.}{./Pictures/}}

\newtheorem{lem}{Lemma}

\begin{document}

\title{Retouched Bloom Filters:\\
Allowing Networked Applications to Trade Off\\
Selected False Positives Against False Negatives}

\author{Benoit Donnet$^{\dag}$, Bruno Baynat$^{\ast}$, Timur Friedman$^{\ast}$\\
$\dag$ Universit\'e Catholique de Louvain, CSE Department\\
$\ast$ Universit\'e Pierre et Marie Curie, Laboratoire LIP6--CNRS
}

\maketitle

\begin{abstract}
Where distributed agents must share voluminous set membership information, 
Bloom filters provide a compact, though lossy, way for them to do so.  Numerous 
recent networking papers have examined the trade-offs between the bandwidth 
consumed by the transmission of Bloom filters, and the error rate, which takes 
the form of false positives, and which rises the more the filters are 
compressed. In this paper, we introduce the retouched Bloom filter (RBF), an 
extension that makes the Bloom filter more flexible by permitting the removal 
of selected false positives at the expense of generating random false 
negatives. We analytically show that RBFs created through a random process 
maintain an overall error rate, expressed as a combination of the false 
positive rate and the false negative rate, that is equal to the false positive 
rate of the corresponding Bloom filters.  We further provide some simple 
heuristics and improved algorithms that decrease the false positive rate more 
than than the corresponding increase in the false negative rate, when creating 
RBFs. Finally, we demonstrate the advantages of an RBF over a Bloom filter in a 
distributed network topology measurement application, where information about 
large stop sets must be shared among route tracing monitors.
\end{abstract}

\section{Introduction}\label{intro}
The \dfn{Bloom filter} is a data structure that was introduced in 
1970~\cite{bloom} and that has been adopted by the networking research 
community in the past decade thanks to the bandwidth efficiencies that it 
offers for the transmission of set membership information between networked 
hosts.  A sender encodes the information into a bit vector, the Bloom filter, 
that is more compact than a conventional representation. Computation and space 
costs for construction are linear in the number of elements.  The receiver uses 
the filter to test whether various elements are members of the set. Though the 
filter will occasionally return a false positive, it will never return a false 
negative. When creating the filter, the sender can choose its desired point in 
a trade-off between the false positive rate and the size. The \dfn{compressed 
Bloom filter}, an extension proposed by Mitzenmacher~\cite{compressed}, allows 
further bandwidth savings.

Broder and Mitzenmacher's survey of Bloom filters' networking 
applications~\cite{survey} attests to the considerable interest in this data 
structure. Variants on the Bloom filter continue to be introduced.  For 
instance, Bonomi et al.'s~\cite{bbf} $d$-left counting Bloom filter is a more 
space-efficient version of Fan et al.'s~\cite{counting} counting Bloom filter, 
which itself goes beyond the standard Bloom filter to allow dynamic insertions 
and deletions of set membership information.  The present paper also introduces 
a variant on the Bloom filter: one that allows an application to remove 
selected false positives from the filter, trading them off against the 
introduction of random false negatives.
  
This paper looks at Bloom filters in the context of a network measurement 
application that must send information concerning large sets of IP addresses 
between measurement points.  Sec.~\ref{case} describes the application in 
detail. But here, we cite two key characteristics of this particular 
application; characteristics that many other networked applications share, and 
that make them candidates for use of the variant that we propose.

First, some false positives might be more troublesome than others, and these 
can be identified after the Bloom filter has been constructed, but before it is 
used.  For instance, when IP addresses arise in measurements, it is not 
uncommon for some addresses to be encountered with much greater frequency than 
others.  If such an address triggers a false positive, the performance 
detriment is greater than if a rarely encountered address does the same. If 
there were a way to remove them from the filter before use, the application 
would benefit.

Second, the application can tolerate a low level of false negatives. It would
benefit from being able to trade off the most troublesome false positives for
some randomly introduced false negatives.

The \dfn{retouched Bloom filter} (RBF) introduced in this paper permits such a 
trade-off.  It allows the removal of selected false positives at the cost of 
introducing random false negatives, and with the benefit of eliminating some 
random false positives at the same time.  An RBF is created from a Bloom filter 
by selectively changing individual bits from 1 to 0, while the size of the 
filter remains unchanged. As Sec.~\ref{rbf.random} shows analytically, an RBF 
created through a random process maintains an overall error rate, expressed as 
a combination of the false positive rate and the false negative rate, that is 
equal to the false positive rate of the corresponding Bloom filter. We further 
provide a number of simple algorithms that lower the false positive rate by a 
greater degree, on average, than the corresponding increase in the false 
negative rate. These algorithms require at most a small constant multiple in 
storage requirements. Any additional processing and storage related to the 
creation of RBFs from Bloom filters are restricted to the measurement points 
that create the RBFs.  There is strictly no addition to the critical resource 
under consideration, which is the bandwidth consumed by communication between 
the measurement points.

Some existing Bloom filter variants do permit the suppression of selected false 
positives, or the removal of information in general, or a trade-off between the 
false positive rate and the false negative rate.  However, as 
Sec.~\ref{related} describes, the RBF is unique in doing so while maintaining 
the size of the original Bloom filter and lowering the overall error rate as 
compared to that filter.

The remainder of this paper is organized as follows: Sec.~\ref{bf} presents the
standard Bloom filter; Sec.~\ref{rbf} presents the RBF, and
shows analytically that the reduction in the false positive rate is equal, on
average, to the increase in the false negative rate even as random 1s in a
Bloom filter are reset to 0s; Sec.~\ref{sc} presents several
simple methods for selectively clearing 1s that are associated with false
positives, and shows through simulations that they reduce the false positive
rate by more, on average, than they increase the false negative rate;
Sec.~\ref{case} describes the use of RBFs in a network measurement application;
Sec.~\ref{related} discusses several Bloom filter variants and compares RBFs to them;
finally, Sec.~\ref{ccl} summarizes the conclusions and future
directions for this work.

\section{Bloom Filters}\label{bf}
A \dfn{Bloom filter} \cite{bloom} is a vector $v$ of $m$ bits that codes the
membership of a subset $A=\{a_1, a_2, \ldots, a_n\}$ of $n$ elements of a
universe $U$ consisting of $N$ elements. In most papers, the size of the
universe is not specified. However, Bloom filters are only useful if the size
of $U$ is much bigger than the size of $A$.

\begin{figure}[!t]
  \begin{center}
    \includegraphics[width=3.5cm]{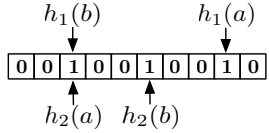}
  \end{center}
\vspace{-0.6cm}
  \caption{A Bloom filter with two hash functions}
  \label{bf.fig}
\end{figure}

The idea is to initialize this vector $v$ to 0, and then take a set $H=\{h_1, 
h_2, \ldots, h_k\}$ of $k$ independent hash functions $h_1, h_2, \ldots, h_k$,
each with range $\{1, \ldots, m\}$.  For each element $a \in A$, the bits at
positions $h_1(a), h_2(a),\ldots, h_k(a)$ in $v$ are set to 1. Note that a
particular bit can be set to 1 several times, as illustrated in
Fig.~\ref{bf.fig}.

\begin{figure}[!t]
  \begin{center} 
    \includegraphics[width=4.5cm]{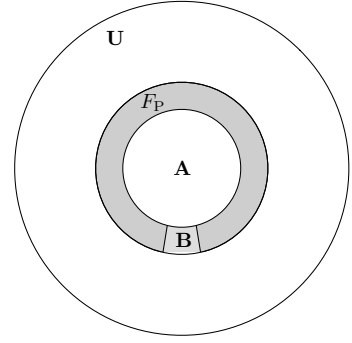}
  \end{center}
  \caption{The false positives set}
  \label{bf.rbf1.fig}
\end{figure}

In order to check if an element $b$ of the universe $U$ belongs to the set $A$,
all one has to do is check that the $k$ bits at positions $h_1(b), h_2(b), 
\ldots, h_k(b)$ are all set to 1. If \textit{at least} one bit is set to 0, we
are sure that $b$ does not belong to $A$. If \textit{all} bits are set to 1, 
$b$ possibly belongs to $A$. There is always a probability that $b$ does not
belong to $A$.  In other words, there is a risk of \dfn{false positives}. Let
us denote by $F_\textrm{P}$ the set of false positives, i.e., the elements that
do not belong to $A$ (and thus that belong to $U-A$) and for which the Bloom
filter gives a positive answer. The sets $U$, $A$, and $F_\textrm{P}$ are
illustrated in Fig.~\ref{bf.rbf1.fig}. ($B$ is a subset of $F_\textrm{P}$ that
will be introduced below.) In Fig.~\ref{bf.rbf1.fig}, $F_\textrm{P}$ is a 
circle surrounding $A$. (Note that $F_\textrm{P}$ is not a superset of $A$.  It
has been colored distinctly to indicate that it is disjoint from $A$.)

We define the \textit{false positive proportion} $f_\textrm{P}$ as the ratio of
the number of elements in $U-A$ that give a positive answer, to the total
number of elements in $U-A$:
\begin{equation}  
  \begin{array}{ll}
        f_\textrm{P} & =~\frac{|F_\textrm{P}|}{|U-A|} 
  \end{array}
  \label{bf.fpprop}
\end{equation}

\begin{figure*}[!t]
  \begin{center}
    \includegraphics[width=5.8cm]{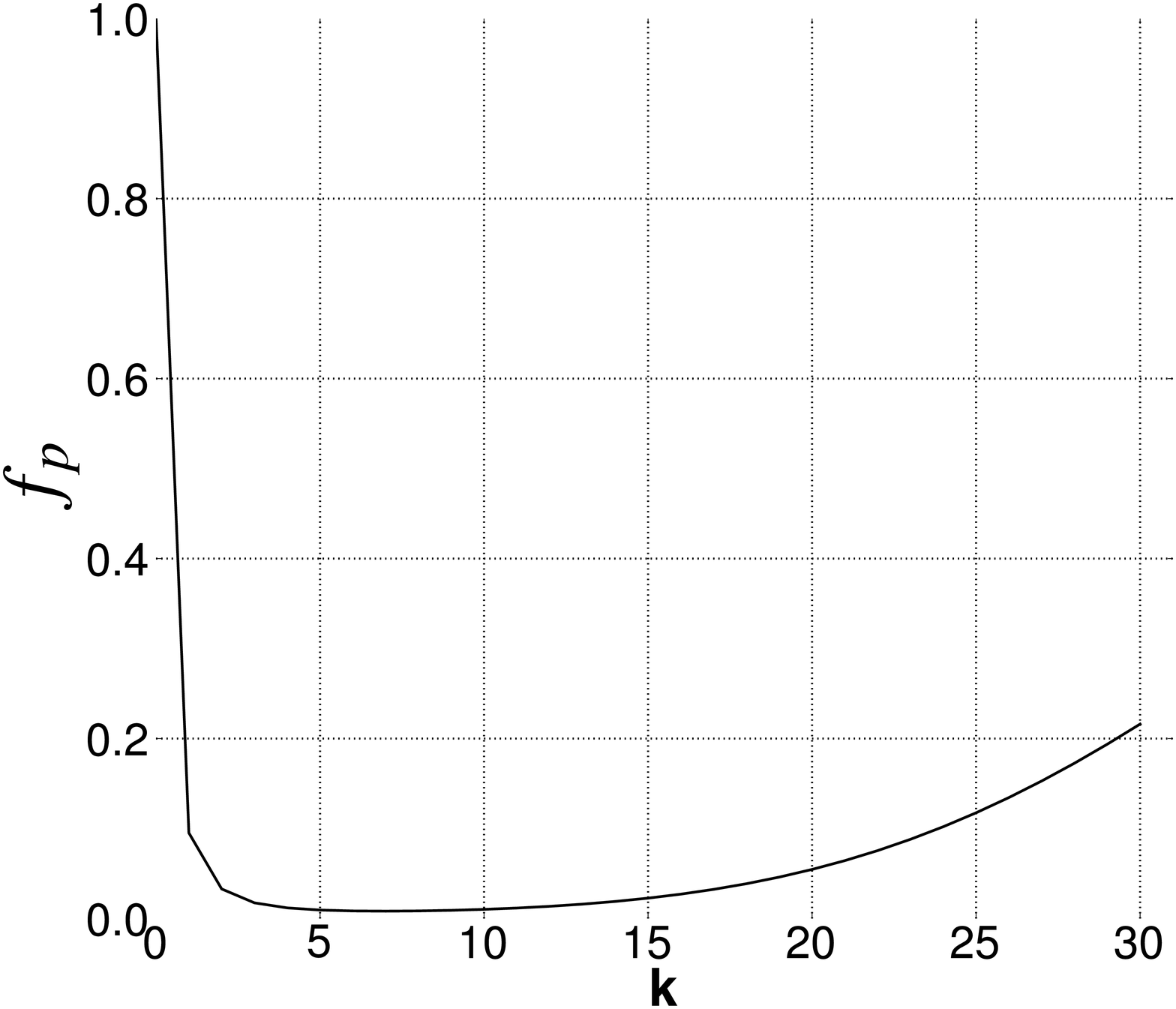}
    \includegraphics[width=5.8cm]{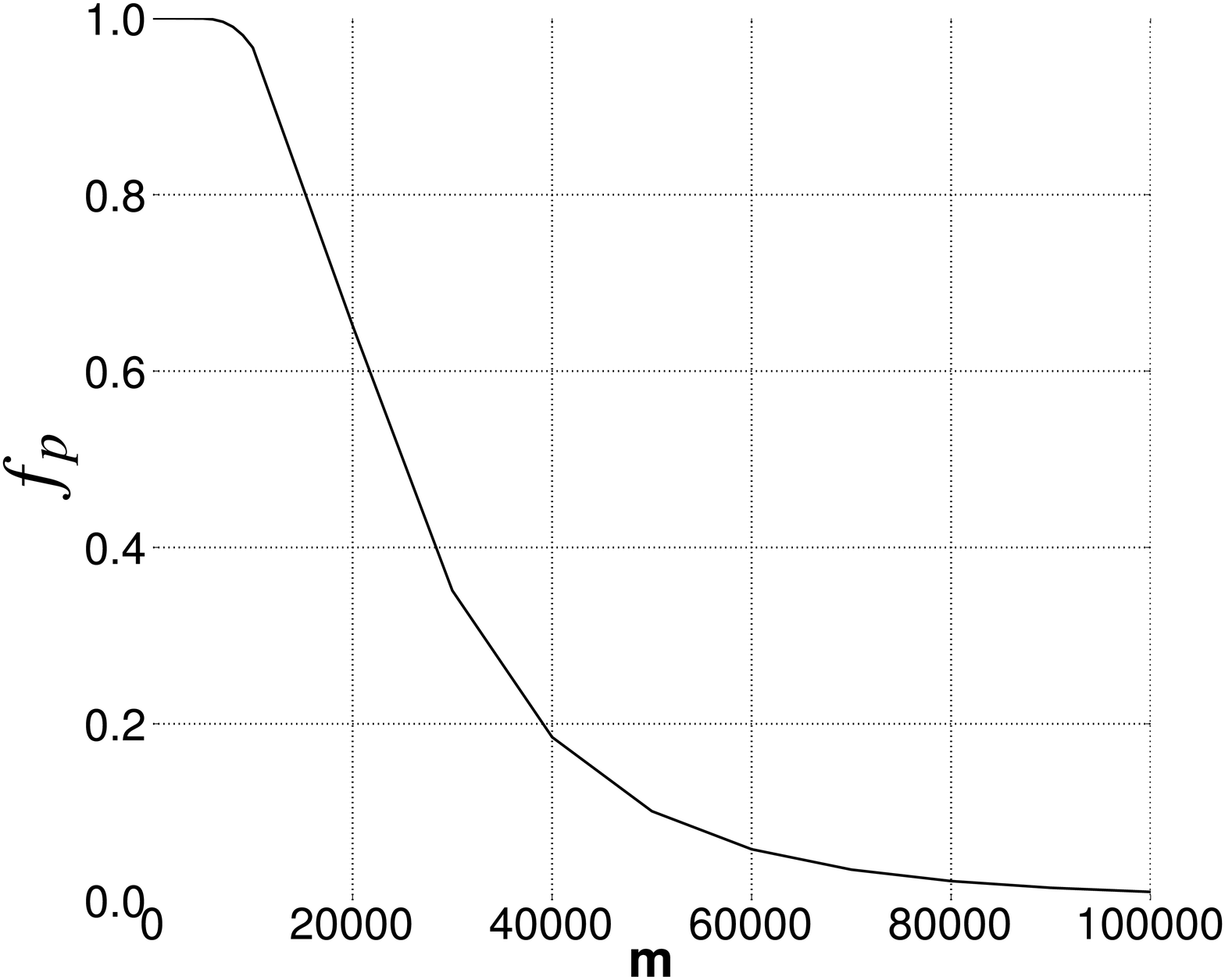}
    \includegraphics[width=5.8cm]{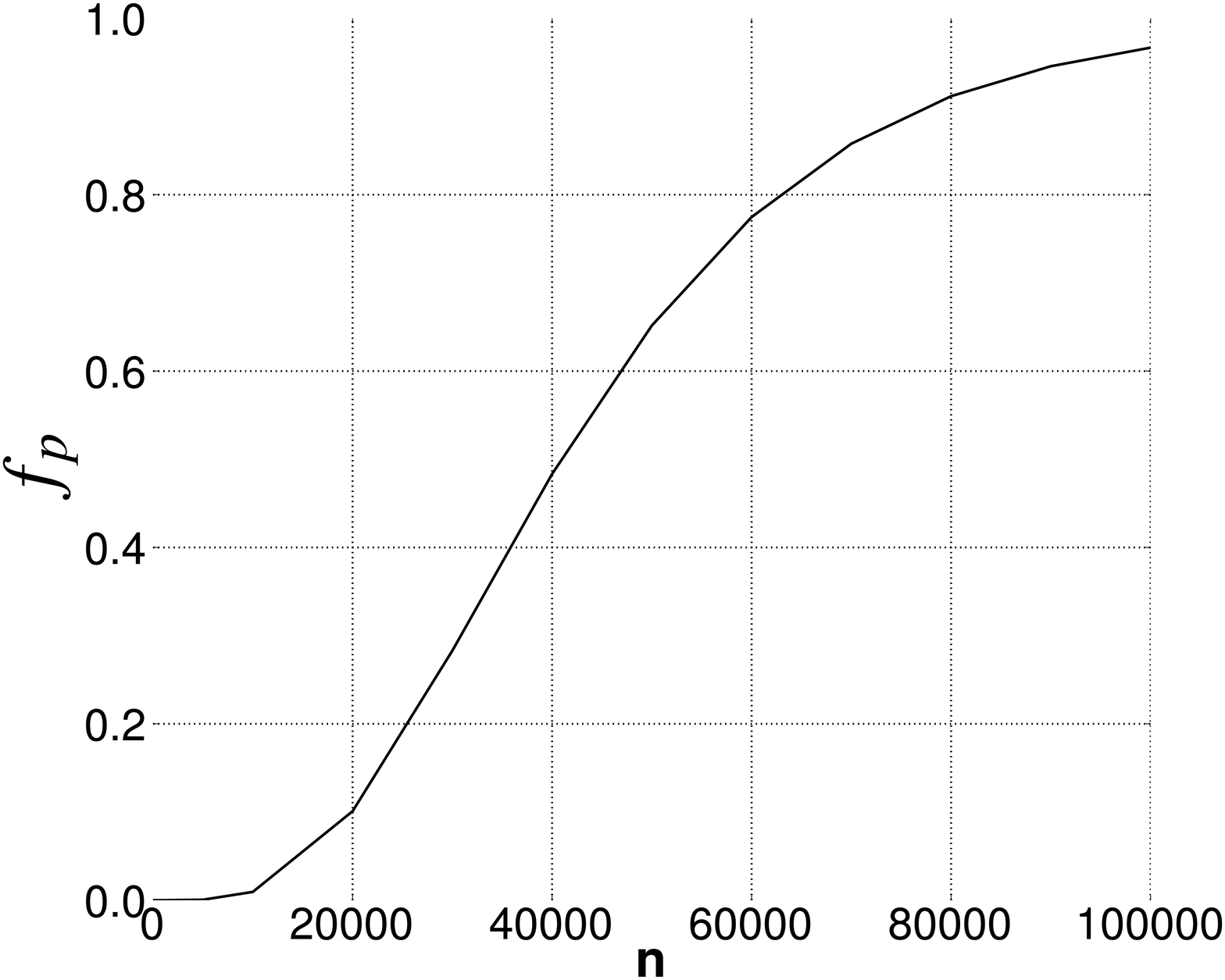}
  \end{center}
  \caption{$f_\textrm{P}$ as a function of $k$, $m$ and $n$.}
  \label{fp(k,m,n).fig}
\end{figure*}

We can alternately define the \dfn{false positive rate}, as the probability
that, for a given element that does not belong to the set $A$, the Bloom filter
erroneously claims that the element is in the set. Note that if this
probability exists (a hypothesis related to the ergodicity of the system that we
assume here), it has the same value as the false positive proportion
$f_\textrm{P}$. As a consequence, we will use the same notation for both
parameters and also denote by $f_\textrm{P}$ the false positive rate. In order
to calculate the false positive rate, most papers assume that all hash
functions map each item in the universe to a random number uniformly over the
range $\{1, \ldots, m\}$. As a consequence, the probability that a specific bit
is set to 1 after the application of one hash function to one element of $A$ is
$\frac{1}{m}$ and the probability that this specific bit is left to 0 is
$1-\frac{1}{m}$. After all elements of $A$ are coded in the Bloom filter, the
probability that a specific bit is always equal to 0 is
\begin{equation}
        p_0~=~\left(1-\frac{1}{m}\right)^{kn}
\label{bf.bit0}
\end{equation}

As $m$ becomes large, $\frac{1}{m}$ is close to zero and $p_0$ can be approximated by
\begin{equation}
        p_0~\approx~e^{-\frac{kn}{m}}
\label{bf.bit0.approx}
\end{equation}

The probability that a specific bit is set to 1 can thus be expressed as
\begin{equation}
  \begin{array}{ll}
p_1 & =~1-p_0
  \end{array}
  \label{bf.bit1}
\end{equation}

The false positive rate can then be estimated by the probability that each of
the $k$ array positions computed by the hash functions is 1. $f_\textrm{P}$ is
then given by
\begin{equation}
  \begin{array}{lll}
        f_\textrm{P} & =~p_1^k\\ & 
        =~\left(1-\left(1-\frac{1}{m}\right)^{kn}\right)^{k}\\ &\approx 
        ~\left(1~-~e^{-\frac{kn}{m}}\right)^{k}
  \end{array}
  \label{bf.fp}
\end{equation}

The false positive rate $f_\textrm{P}$ is thus a function of three parameters:
$n$, the size of subset $A$; $m$, the size of the filter; and $k$, the number of
hash functions.  Fig.~\ref{fp(k,m,n).fig} illustrates the variation of
$f_\textrm{P}$ with respect to the three parameters individually (when the two
others are held constant). Obviously, and as can be seen on these graphs,
$f_\textrm{P}$ is a decreasing function of $m$ and an increasing function of
$n$.  Now, when $k$ varies (with $n$ and $m$ constant), $f_\textrm{P}$ first
decreases, reaches a minimum and then increases. Indeed there are two
contradicting factors: using more hash functions gives us more chances to find
a $0$ bit for an element that is not a member of $A$, but using fewer hash
functions increases the fraction of $0$ bits in the array. As stated, e.g., by
Fan et al.~\cite{counting}, $f_\textrm{P}$ is minimized when
\begin{equation}
k~=~\frac{m \ln 2}{n}
\label{bf.nbHash}
\end{equation}
for fixed $m$ and $n$. Indeed, the derivative of $f_\textrm{P}$ (estimated by
eqn.~\ref{bf.bit0.approx}) with respect to $k$ is $0$ when $k$ is given by
eqn.~\ref{bf.nbHash}, and it can further be shown that this is a global minimum.

Thus the minimum possible false positive rate for given values of $m$ and $n$
is given by eqn.~\ref{bf.minFPR}. In practice, of course, $k$ must be an
integer. As a consequence, the value furnished by eqn.~\ref{bf.nbHash} is rounded to the
nearest integer and the resulting false positive rate will be somewhat higher than
the optimal value given in eqn.~\ref{bf.minFPR}.
\begin{equation}
        \hat{f}_\textrm{P}~=~\left(\frac{1}{2}\right)^ {\frac{m \ln 2}{n}}
        \approx~\left(0.6185\right)^ {\frac{m}{n}}
\label{bf.minFPR}
\end{equation}

Finally, it is important to emphasize that the absolute number of false
positives is relative to the  size of $U-A$ (and not directly to the size of
$A$). This result seems surprising as the expression of $f_\textrm{P}$ depends
on $n$, the size of $A$, and does not depend on $N$, the size of $U$. If we
double the size of $U-A$ (and keep the size of $A$ constant) we also double the
absolute number of false positives (and obviously the false positive rate is
unchanged).

\section{Retouched Bloom Filters}\label{rbf}
As shown in Sec.~\ref{bf}, there is a trade-off between the size of the Bloom
filter and the probability of a false positive. For a given $n$, even by
optimally choosing the number of hash functions, the only way to reduce the
false positive rate in standard Bloom filters is to increase the size $m$ of
the bit vector. Unfortunately, although this implies a gain in terms of a reduced
false positive rate, it also implies a loss in terms of increased memory usage.
Bandwidth usage becomes a constraint that must be minimized when Bloom filters
are transmitted in the network.

\subsection{Bit Clearing}\label{rbf.bc}
In this paper, we introduce an extension to the Bloom filter, referred to as
the \dfn{retouched Bloom filter} (RBF).  The RBF makes standard Bloom filters
more flexible by allowing selected false positives to be traded off against
random false negatives.  False negatives do not arise at all in the standard
case. The idea behind the RBF is to remove a certain number of these selected
false positives by resetting individually chosen bits in the vector $v$. We
call this process the \dfn{bit clearing process}. Resetting a given bit to 0
not only has the effect of removing a certain number of false positives,  but
also generates false negatives. Indeed, any element $a \in A$ such that (at
least) one of the $k$ bits at positions $h_1(a), h_2(a), \ldots, h_k(a)$ has
been reset to 0, now triggers a negative answer. Element $a$ thus becomes a
false negative.

\begin{figure}[!t]
  \begin{center}
    \includegraphics[width=4.5cm]{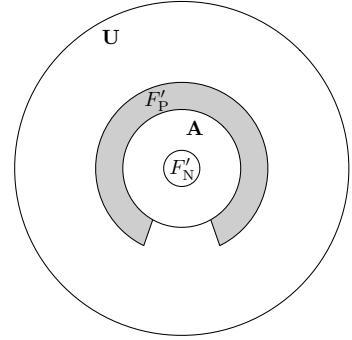}
  \end{center}
  \caption{False positive and false negative sets after the selective clearing process}
  \label{rb.rbf2.fig}
\end{figure}

To summarize, the bit clearing process has the effects of decreasing the number
of false positives and of generating a number of false negatives. Let us use
the labels $F_\textrm{P}'$ and $F_\textrm{N}'$ to describe the sets of false
positives and false negatives after the bit clearing process. The sets
$F_\textrm{P}'$ and $F_\textrm{N}'$ are illustrated in Fig.~\ref{rb.rbf2.fig}.

After the bit clearing process, the false positive and false negative
proportions are given by
\begin{equation}
  \begin{array}{ll}
        f_\textrm{P}' & =~\frac{|F_\textrm{P}'|}{|U-A|}
  \end{array}
  \label{bf.fppropaftersc}
\end{equation}
\begin{equation}
  \begin{array}{ll}
        f_\textrm{N}' & =~\frac{|F_\textrm{N}'|}{|A|}
  \end{array}
  \label{bf.fnpropaftersc}
\end{equation}

Obviously, the false positive proportion has decreased (as $F_\textrm{P}'$ is
smaller than $F_\textrm{P}$) and the false negative proportion has increased
(as it was zero before the clearing). We can measure the benefit of the bit
clearing process by introducing $\Delta f_\textrm{P}$, the proportion of false
positives removed by the bit clearing process, and $\Delta f_\textrm{N}$, the
proportion of false negatives generated by the bit clearing process:
\begin{equation}
\Delta 
f_\textrm{P}~=~\frac{|F_\textrm{P}|-|F_\textrm{P}'|}{|F_\textrm{P}|}=\frac{f_\textrm{P}-f_\textrm{P}'}{f_\textrm{P}}
  \label{bf.deltafp}
\end{equation}
\begin{equation}
        \Delta f_\textrm{N}~=~\frac{|F_\textrm{N}'|}{|A|}=f_\textrm{N}'
\label{bf.deltafn}
\end{equation}

We, finally, define $\chi$ as the ratio between the proportion of false
positives removed and the proportion of false negatives generated:
\begin{equation}
        \chi~=~\frac{\Delta f_\textrm{P}}{\Delta f_\textrm{N}}
  \label{bf.chi}
\end{equation}

$\chi$ is the main metric we introduce in this paper in order to evaluate the
RBF.  If $\chi$ is greater than 1, it means that the proportion of false
positives removed is higher than the proportion of false negatives generated.

\subsection{Randomized Bit Clearing}\label{rbf.random}
In this section, we analytically study the effect of randomly resetting bits in
the Bloom filter, whether these bits correspond to false positives or not. We
call this process the \textit{randomized bit clearing process}. In
Sec.~\ref{sc}, we discuss more sophisticated approaches to choosing the 
bits that should be cleared. However, performing random clearing in the Bloom
filter enables us to derive analytical results concerning the consequences of
the clearing process. In addition to providing a formal derivation of the
benefit of RBFs, it also gives a lower bound on the
performance of any smarter selective clearing approach (such as those developed
in Sec.~\ref{sc}).

We again assume that all hash functions map each element of the universe $U$ to
a random number uniformly over the range $\{1, \ldots, m\}$. Once the $n$ 
elements of $A$ have been coded in the Bloom filter, there is a probability
$p_0$ for a given bit in $v$ to be $0$ and a probability $p_1$ for it to be 
$1$. As a consequence, there is an average number of $p_1m$ bits set to $1$ in
$v$. Let us study the effect of resetting to $0$ a randomly chosen bit in $v$. 
Each of the $p_1m$ bits set to $1$ in $v$ has a probability $\frac{1}{p_1m}$ of
being reset and a probability $1-\frac{1}{p_1m}$ of being left at $1$.

The first consequence of resetting a bit to $0$ is to remove a certain number
of false positives. If we consider a given false positive $x \in F_\textrm{P}$,
after the reset it will not result in a positive test any more if the bit that
has been reset belongs to one of the $k$ positions $h_1(x), h_2(x), \ldots, 
h_k(x)$. Conversely, if none of the $k$ positions have been reset, $x$ remains
a false positive. The probability of this latter event is
\begin{equation}
r_1~=~\left(1-\frac{1}{p_1m}\right)^k
  \label{bf.r1}
\end{equation}

As a consequence, after the reset of one bit in $v$, the false positive rate
decreases from  $f_\textrm{P}$ (given by eqn.~\ref{bf.fp}) to
$f_\textrm{P}'=f_\textrm{P}r_\textrm{1}$. The proportion of false positives
that have been eliminated by the resetting of a randomly chosen bit in $v$ is
thus equal to $1-r_1$:
\begin{equation}
        \Delta f_\textrm{P}~=~1-r_1
  \label{bf.deltafp2}
\end{equation}

The second consequence of resetting a bit to $0$ is the generation of a certain
number of false  negatives. If we consider a given element $a \in A$, after the
reset it will result in a negative test if the bit that has been reset in $v$
belongs to one of the $k$ positions $h_1(a), h_2(a), \ldots, h_k(a)$.
Conversely, if none of the $k$ positions have been reset, the test on $a$
remains positive. Obviously, the probability that a given element in $A$
becomes a false negative is given by $1-r_1$  (the same reasoning holds):
\begin{equation}
\Delta f_\textrm{N}~=~1-r_1
  \label{bf.deltafn2}
\end{equation}

We have demonstrated that resetting one bit to $0$ in $v$ has the effect of 
eliminating  the same proportion of false positives as the proportion of false
negatives generated. As a result, $\chi=1$.  It is however important to note 
that the proportion of false positives that are eliminated is relative to the
size of the set of false positives (which in turns is relative to the size of
$U-A$, thanks to eqn.~\ref{bf.fp}) whereas the proportion of false negatives
generated is relative to the size of $A$. As we assume that $U-A$ is much 
bigger than $A$ (actually if $|F_\textrm{P}| > |A|$), resetting a bit to $0$ in
$v$ can eliminate many more false positives than the number of false negatives
generated.

It is easy to extend the demonstration to the reset of $s$ bits and see that it
eliminates a  proportion $1-r_s$ of false positives and generates the same
proportion of false negatives, where $r_s$ is given by
\begin{equation}
        r_s~=~\left(1-\frac{s}{p_1m}\right)^k
  \label{bf.r2}
\end{equation}

As a consequence, any random clearing of bits in the Bloom vector $v$ has the
effect of maintaining the ratio $\chi$ equal to $1$.

\section{Selective Clearing}\label{sc}
Sec.~\ref{rbf} introduced the idea of randomized bit clearing and analytically
studied the effect of randomly resetting $s$ bits of $v$, whether these bits
correspond to false positives or not.  We showed that it has the effect of
maintaining the ratio $\chi$ equal to 1. In this section, we refine the idea of
randomized bit clearing by focusing on bits corresponding to elements that trigger
false positives. We call this process \dfn{selective clearing}.

As described in Sec.~\ref{bf}, in Bloom filters (and also in RBFs),
some elements in $U - A$ will trigger false positives, forming the set
$F_\textrm{P}$. However, in practice, it is likely that not all false positives
will be encountered.  To illustrate this assertion, let us assume that the
universe $U$ consists of the whole IPv4 addresses range.  To build the Bloom
filter or the RBF, we define $k$ hash functions based on a
32 bit string. The subset $A$ to record in the filter is a small portion of the
IPv4 address range.  Not all false positives will be encountered in practice
because a significant portion of the IPv4 addresses in $F_\textrm{P}$ have not
been assigned.

We record the false positives encountered in practice in a set called $B$, with 
$B \subseteq F_\textrm{P}$ (see Fig.~\ref{bf.rbf1.fig}). Elements in $B$ are 
false positives that we label as \dfn{troublesome keys}, as they generate, when 
presented as keys to the Bloom filter's hash functions, false positives that 
are liable to be encountered in practice.  We would like to eliminate the 
elements of $B$ from the filter.

In the following sections, we explore several algorithms for performing
selective clearing (Sec.~\ref{sc.algo}).  We then evaluate and compare the
performance of these algorithms using theorical analysis (Sec.~\ref{sc.theor})
and simulation analysis (Sec.~\ref{sc.simu}).

\subsection{Algorithms}\label{sc.algo}
In this section, we propose four different algorithms that allow one to remove 
the false positives belonging to $B$.  All of these algorithms are simple to 
implement and deploy.  We first present an algorithm that does not require any 
intelligence in selective clearing.  Next, we propose refined algorithms that 
take into account the risk of false negatives.  With these algorithms, we show 
how to trade-off false positives for false negatives.

The first algorithm is called \dfn{Random Selection}.  The main idea is, for
each troublesome key to remove, to randomly select a bit amongst the $k$
available to reset. The main interest of the Random Selection algorithm is its
extreme computational simplicity: no effort has to go into selecting a bit to
clear.  Random Selection differs from random clearing (see Sec.~\ref{rbf}) by
focusing on a set of troublesome keys to remove, $B$, and not by resetting
randomly any bit in $v$, whether it corresponds to a false positive or not.
Random Selection is formally defined in Algorithm~\ref{sc.algo.rs}.

\begin{algorithm}[!t]
  \caption{Random Selection}
  \label{sc.algo.rs}
  \begin{algorithmic}[1]
    \Require $v$, the bit vector.
    \Ensure $v$ updated, if needed.
    \Procedure{RandomSelection}{$B$}
      \ForAll{$b_i \in B$}
        \If{\textsc{MembershipTest}($b_i$, $v$)}
          \State index $\leftarrow$ \textsc{Random}($h_1(b_i)$, \ldots, $h_k(b_i)$)
          \State $v$[index] $\leftarrow$ 0
        \EndIf
      \EndFor
    \EndProcedure
  \end{algorithmic}
\end{algorithm}

Recall that $B$ is the set of troublesome keys to remove.   This set can 
contain from only one element to the whole set of false positives. Before 
removing a false positive element, we make sure that this element is still 
falsely recorded in the RBF, as it could have been removed previously.  Indeed, 
due to collisions that may occur between hashed keys in the bit vector, as 
shown in Fig.~\ref{bf.fig}, one of the $k$ hashed bit positions of the element 
to remove may have been previously reset. Algorithm~\ref{sc.algo.rs} assumes 
that a function \textsc{Random} is defined and returns a value randomly chosen 
amongst its uniformly distributed  arguments. The algorithm also assumes that 
the function \textsc{MembershipTest} is defined.  It takes two arguments: the 
key to be tested and the bit vector.  This function returns \textit{true} if 
the element is recorded in the bit vector (i.e., all the $k$ positions 
corresponding to the hash functions are set to 1). It returns \textit{false} 
otherwise.

The second algorithm we propose is called \dfn{Minimum FN Selection}. The idea 
is to minimize the false negatives generated by each selective clearing. For 
each troublesome key to remove that was not previously cleared, we choose 
amongst the $k$ bit positions the one that we estimate will generate the 
minimum number of false negatives.   This minimum is given by the 
\textsc{MinIndex} procedure in Algorithm~\ref{sc.algo.mfn}.  This can be 
achieved by maintaining locally a counting vector, $v_A$, storing in each 
vector position the quantity of elements recorded. This algorithm effectively 
takes into account the possibility of collisions in the bit vector between 
hashed keys of elements belonging to $A$.  Minimum FN Selection is formally 
defined in Algorithm~\ref{sc.algo.mfn}.

For purposes of algorithmic simplicity, we do not entirely update the counting 
vector with each iteration.  The cost comes in terms of an over-estimation, for 
the heuristic, in assessing the number of false negatives that it introduces in 
any given iteration.  This over-estimation grows as the algorithm progresses. 
We are currently studying ways to efficiently adjust for this over-estimation. 
Sec.~\ref{isc} will discuss more complex selective clearing algorithms that 
update, at each step, the counting vector.

\begin{algorithm}[!t]
  \caption{Minimum FN Selection}
  \label{sc.algo.mfn}
  \begin{algorithmic}[1]
    \Require $v$, the bit vector and $v_A$, the counting vector.
    \Ensure $v$ and $v_A$ updated, if needed.
    \Procedure{MinimumFNSelection}{$B$}
      \State \textsc{CreateCV}($A$)
      \ForAll{$b_i \in B$}
        \If{\textsc{MembershipTest}($b_i$, $v$)}
          \State index $\leftarrow$ \textsc{MinIndex}($b_i$)
          \State $v$[index] $\leftarrow$ 0
          \State $v_A$[index] $\leftarrow$ 0
        \EndIf
      \EndFor
    \EndProcedure
    \State
     \Procedure{CreateCV}{$A$}
      \ForAll{$a_i \in A$}
        \For{$j=1 \mathrm{\ to\ } k$}
          \State $v_A$[$h_j(a_i)$]++
        \EndFor
      \EndFor
    \EndProcedure
  \end{algorithmic}
\end{algorithm}

The third selective clearing mechanism is called \dfn{Maximum FP Selection}. In 
this case,  we try to maximize the quantity of false positives to remove. For 
each troublesome key to remove that was not previously deleted, we choose 
amongst the $k$ bit positions the one we estimate to allow removal of the 
largest number of false positives, the position of which is given by the 
\textsc{MaxIndex} function in Algorithm~\ref{sc.algo.mfp}.  In the fashion of 
the Minimum FN Selection algorithm, this is achieved by maintaining a counting 
vector, $v_B$, storing in each vector position the quantity of false positive 
elements recorded. For each false positive element, we choose the bit 
corresponding to the largest number of false positives recorded. This algorithm 
considers as an opportunity the risk of collisions in the bit vector between 
hashed keys of elements generating false positives.   Maximum FP Selection is 
formally described in Algorithm~\ref{sc.algo.mfp}.

\begin{algorithm}[!t]
  \caption{Maximum FP Selection}
  \label{sc.algo.mfp}
  \begin{algorithmic}[1]
    \Require $v$, the bit vector and $v_B$, the counting vector.
    \Ensure $v$ and $v_B$ updated, if needed.
    \Procedure{MaximumFP}{$B$}
      \State \textsc{CreateFV}($B$)
      \ForAll{$b_i \in B$}
        \If{\textsc{MembershipTest}($b_i$, $v$)}
          \State index $\leftarrow$ \textsc{MaxIndex}($b_i$)
          \State $v$[index] $\leftarrow$ 0
          \State $v_B$[index] $\leftarrow$ 0
        \EndIf
      \EndFor
    \EndProcedure
    \State
    \Procedure{CreateFV}{$B$}
      \ForAll{$b_i \in B$}
        \For{$j=1 \mathrm{\ to\ } k$}
          \State $v_B$[$h_j(b_i)$]++
        \EndFor
      \EndFor
    \EndProcedure
  \end{algorithmic}
\end{algorithm}

Finally, we propose a selective clearing mechanism called \dfn{Ratio 
Selection}.  The idea is to combine Minimum FN Selection and Maximum FP 
Selection into a single algorithm.  Ratio Selection provides an approach in 
which we try to minimize the false negatives generated while maximizing the 
false positives removed.  Ratio Selection therefore takes into account the risk 
of collision between hashed keys of elements belonging to $A$ and hashed keys 
of elements belonging to $B$. It is achieved by maintaining a ratio vector, 
$r$, in which each position is the ratio between $v_A$ and $v_B$. For each 
troublesome key that was not previously cleared, we choose the index where the 
ratio is the minimum amongst the $k$ ones.  This index is given by the 
\textsc{MinRatio} function in Algorithm~\ref{sc.algo.ratio}. Ratio Selection is 
defined in Algorithm~\ref{sc.algo.ratio}.  This algorithm makes use of the 
\textsc{CreateCV} and \textsc{CreateFV} functions previously defined for 
Algorithms~\ref{sc.algo.mfn} and \ref{sc.algo.mfp}.

\begin{algorithm}[!t]
  \caption{Ratio Selection}
  \label{sc.algo.ratio}
  \begin{algorithmic}[1]
    \Require $v$, the bit vector, $v_B$ and $v_A$, the counting
      vectors and $r$, the ratio vector.
    \Ensure $v$, $v_A$, $v_B$ and $r$ updated, if needed.
    \Procedure{Ratio}{$B$}
      \State \textsc{CreateCV}($A$)
      \State \textsc{CreateFV}($B$)
      \State \textsc{ComputeRatio}()
      \ForAll{$b_i \in B$}
        \If{\textsc{MembershipTest}($b_i$, $v$)}
                  \State index $\leftarrow$ \textsc{MinRatio}($b_i$)
          \State $v$[index] $\leftarrow$ 0
          \State $v_A$[index] $\leftarrow$ 0
          \State $v_B$[index] $\leftarrow$ 0
          \State $r$[index] $\leftarrow$ 0
        \EndIf
      \EndFor
    \EndProcedure
    \State
    \Procedure{ComputeRatio}{}
      \For{$i=1 \mathrm{\ to\ } m$}
        \If{$v$[i] $\wedge$ $v_B[i] > 0$}
          \State $r$[i] $\leftarrow$ $\frac{v_A[i]}{v_B[i]}$
        \EndIf
      \EndFor
    \EndProcedure
  \end{algorithmic}
\end{algorithm}

\subsection{Theorical Analysis}\label{sc.theor}
\subsubsection{Algorithmic Complexity}\label{sc.theor.alg}
\begin{lem}
The algorithmic complexity of the Random Selection algorithm is $O(k \times 
|B|)$.
\end{lem}
\begin{proof}
Before going into details of the Random Selection algorithm, let us first have
a look at the \textsc{MembershipTest} procedure.  This procedure takes two
arguments: $x_i$, an element belonging to $|B|$ and $v$, the bit vector.  The
\textsc{MembershipTest} procedure aims at determining whether the element $x_i$
is recorded in the bit vector $v$, or not.  Therefore, as explained in
Sec.~\ref{bf}, the \textsc{MembershipTest} procedure checks if the $k$ bits at
positions $h_1(b), h_2(b), \ldots, h_k(b)$ are all set to 1.  As a consequence,
the algorithmic complexity of the \textsc{MembershipTest} is $O(k)$.  

Now, let us consider the Random Selection algorithm in its entirety. Random 
Selection browses all elements belonging to $B$.  And for each element in $B$, 
Random Selection calls the \textsc{MembershipTest} procedure. Therefore, the 
\textsc{MembershipTest} procedure is called $|B|$ times. 

Consequently, the algorithmic complexity of the Random Selection is $O(k \times
|B|)$. 
\end{proof}

\begin{lem}
The running time of the Minimum FN Selection algorithm is $O(k \times (|A| + |B|))$.
\end{lem}
\begin{proof}
We first have a look at the \textsc{CreateCV} procedure.  \textsc{CreateCV} aims
at creating the counting vector $v_A$ that indicates, for each cell, the number
of element recorded in the corresponding cell of the bit vector $v$.  Therefore,
this procedure browses all elements belonging to $A$ and, for each element,
increments $k$ counters, where $k$ gives the number of hash functions used. 
Consequently, the algorithmic complexity of the \textsc{CreateCV} procedure is
$O(k \times |A|)$.  

After returning from the \textsc{CreateCV} procedure call, the Minimum FN 
Selection algorithm browses all elements belonging to $B$ and, for each 
element, calls \textsc{MembershipTest}.  If the membership test returns true, 
then the \textsc{MinIndex} procedure is called. This procedure aims at 
determining the bit vector index that returns the minimum value among $k$ 
available.  The algorithmic complexity is thus $O(k)$.  

Until now, the complexity of Minimum FN Selection is $O(\max(k \times |A|, 2k 
\times |B|))$.  The term $2k$ can be reduced to $k$.  Finally, it is easy to 
show that $O(k \times \max(|A|, |B|))$ is equivalent to $O(k \times (|A| + 
|B|))$.
\end{proof}

\begin{lem}
The running time of the Maximum FP Selection algorithm is $O(k \times |B|)$.
\end{lem}
\begin{proof}
Let us first consider the \textsc{CreateFV} procedure.  This aims at creating
the counting vector $v_B$ that indicates, for each cell, the number of false
positives recorded in the corresponding cell of the bit vector $v$.  Therefore,
this procedure browses all elements belonging to $B$ and, for each element,
increments $k$ counters, where $k$ gives the number of hash functions used. 
Consequently, the algorithmic complexity of the \textsc{CreateFV} procedure is
$O(k \times |B|)$.  

After returning from the \textsc{CreateFV} procedure call, the Maximum FP 
Selection algorithm browses all elements belonging to $B$ and, for each 
element, calls \textsc{MembershipTest}.  If the membership test returns true, 
then the \textsc{MaxIndex} procedure is called. This procedure aims at 
determining the bit vector index that returns the maximum value among $k$ 
available.  The algorithmic complexity is thus $O(k)$.  

Until now, the complexity of Maximum FP Selection is $O(2 \times (k \times 
|B|))$.  The multiplicative factor $2$ is negligible.  Consequently, the 
algorithmic complexity of the Maximum FP Selection algorithm is 
$O(k \times |B|)$.
\end{proof}

\begin{lem}
The running time of the Ratio Selection algorithm is $O(k \times (|A| + |B|) + m)$.
\end{lem}
\begin{proof}
As explained above, the complexity of \textsc{CreateCV} is $O(k \times |A|)$
and \textsc{CreateFV} is $O(k \times |B|)$.  After calling \textsc{CreateCV} and
\textsc{CreateFV}, the Ratio Selection algorithm calls the \textsc{Ratio}
procedure that aims at creating the ratio of $v_A$ to $v_B$.  The complexity of
\textsc{Ratio} is $O(m)$ as it must browses all vector cells.  

The rest of Ratio Selection behaves the same way as Minimum FN Selection and 
Maximum FP Selection, i.e., it browses all elements belonging to $B$, performs 
the membership test and, if needed, selects the minimum value among $k$ 
available. Therefore, the complexity is $O(k \times (|A| + |B|))$ to which we 
add the cost associated to the \textsc{ComputeRatio} procedure, i.e. $O(m)$.

Consequently, the algorithmic complexity is $O(k \times (|A| + 
|B|) + m)$.
\end{proof}

\subsubsection{Spatial Complexity}\label{sc.theor.spat}
\begin{lem}
The spatial complexity of the Random Selection algorithm is $O(m + |B|)$
\end{lem}
\begin{proof}
The Random Selection algorithm makes use of two data structures: $v$, the bit
vector required by the Bloom filters, and $B$, the set of troublesome keys to
remove from the Bloom filter.  The vector $v$ is $m$ bit long.  Therefore, the
spatial complexity of the Random Selection algorithm is $O(m + |B|)$.
\end{proof}

\begin{lem}
The spatial complexity of the Minimum FN Selection algorithm is $O(cm + |B|)$
\end{lem}
\begin{proof}
The Minimum FN Selection algorithm makes use of three data structures: $v$, the
$m$ bit vector, $B$, the set of troublesome keys to remove from the Bloom filter
and $v_A$, the counting vector.  $v_A$ is $m$ cells long and each cell contains
$c$ bits needed by the counter.  Therefore, the spatial complexity of the
Minimum FN Selection algorithm is $O(cm + |B|)$.
\end{proof}

\begin{lem}
The spatial complexity of the Maximum FP Selection algorithm is $O(cm + |B|)$.
\end{lem}
\begin{proof}
The Maximum FP Selection algorithm makes use of three data structures: $v$, the
$m$ bit vector, $B$, the set of troublesome keys to remove from the Bloom filter
and $v_B$, the counting vector.  $v_B$ is $m$ cells long and each cell contains
$c$ bits needed by the counter.  Therefore, the spatial complexity of the
Maximum FP Selection algorithm is $O(cm + |B|)$.
\end{proof}

\begin{lem}
The spatial complexity of the Ratio Selection algorithm is $O(cm + dm + |B|)$. 
\end{lem}
\begin{proof}
The Ratio Selection algorithm makes use of four data structures: $v$, the $m$
bit vector, $B$, the set of troublesome keys to remove from the Bloom filter,
$v_A$, the counting vector of elements truly recorded in $v$, $v_B$, the
counting vector of false positives recorded in $v$ and $r$, the ratio vector. 
$r$ is $m$ cells long and each cell contains $d$ bits needed by the counter. 
Note that $d$ is greater than $c$  as $r$ records ratios.  Therefore, the
spatial complexity Ratio algorithm is $O(cm + dm + |B|)$.
\end{proof}

\subsection{Simulation Analysis}\label{sc.simu}
\subsubsection{Methodology}\label{sc.simu.methodo}
We conducted an experiment with a universe $U$ of 2,000,000 elements ($N = 
2,000,000$). These elements, for the sake of simplicity, were integers 
belonging to the range [0; 1,999,9999].  The subset $A$ that we wanted to 
summarize in the Bloom filter contains 10,000 different elements ($n = 
10,000$) randomly chosen from the universe $U$.  Bloom's 
paper~\cite{bloom} states that $|U|$ must be much greater than $|A|$, without 
specifying a precise scale.

The bit vector $v$ we used for simulations is 100,000 bits long ($m = 
100,000$), ten times bigger than $|A|$. The RBF used five different and 
independent hash functions ($k=5$). Hashing was emulated with random numbers.  
We simulated randomness with the Mersenne Twister MT19937 pseudo-random number 
generator~\cite{twister}.  Using five hash functions and a bit vector ten times 
bigger than $n$ is advised by Fan et al.~\cite{counting}. This permits a good 
trade-off between membership query accuracy, i.e., a low false positive rate of 
0.0094 when estimated with eqn.~\ref{bf.fp}, memory usage and computation time. 
As mentioned earlier in this paper (see Sec.~\ref{bf}), the false positive rate 
may be decreased by increasing the bit vector size but it leads to a lower 
compression level.

For our experiment, we defined the ratio of troublesome keys compared to the
entire set of false positives as
\begin{equation}
  \beta~=~\frac{|B|}{|F_P|}
\label{sc.eval.methodo.beta}
\end{equation}

We considered the following values of $\beta$: 1\%, 2\%, 5\%, 10\%, 25\%,
50\%, 75\% and 100\%.  When $\beta = 100$\%, it means that $B = F_\textrm{P}$
and we want to remove all the false positives.

Each data point in the plots and tables represents the mean value over fifteen 
runs of the experiment, each run using a new $A$, $F_\textrm{P}$, $B$, and RBF. 
We determined 95\% confidence intervals for the mean based on the Student $t$ 
distribution.

We performed the experiment as follows: we first created the universe $U$ and 
randomly affected 10,000 of its elements to $A$.  We next built $F_\textrm{P}$ 
by applying the following scheme.  Rather than using eqn.~\ref{bf.fp} to 
compute the false positive rate and then creating $F_\textrm{P}$ by randomly 
affecting positions in $v$ for the false positive elements, we preferred to 
experimentally compute the false positives. We queried the RBF with a 
membership test for each element belonging to $U-A$. False positives were the 
elements that belong to the Bloom filter but not to $A$.  We kept track of them 
in a set called $F_\textrm{P}$. This process seemed to us more realistic 
because we evaluated the real quantity of false positive elements in our data 
set. $B$ was then constructed by randomly selecting a certain quantity of 
elements in $F_\textrm{P}$, the quantity corresponding to the desired 
cardinality of $B$. We next removed all troublesome keys from $B$ by using one 
of the selective clearing algorithms, as explained in Sec.~\ref{sc.algo}.  We 
then built $F_\textrm{N}'$, the false negative set, by testing all elements in 
$A$ and adding to $F_\textrm{N}'$ all elements that no longer belong to $A$. We 
also determined $F_\textrm{P}'$, the false positive set after removing the set 
of troublesome keys $B$.

\subsubsection{Results}\label{sc.simu.res}
\begin{table*}[!t]
  \begin{center}
    \begin{tabular}{c|rrrrrrrr}
         & \multicolumn{2}{c}{$|B|$} & \multicolumn{2}{c}{$|B'|$} & \multicolumn{2}{c}{$|B|+|B'|$} & \multicolumn{2}{c}{$|A'|$}\\
      \hline
     1\%   & \co{188}{1.31}     & \co{434}{13.74}  & \co{622}{13.84}    & \co{231}{3.01}\\
     2\%   & \co{375}{1.84}     & \co{842}{21.84}  & \co{1217}{22.85}   & \co{450}{7.75}\\
     5\%   & \co{932}{9.94}     & \co{1934}{37.83} & \co{2826}{46.21}   & \co{1070}{10.05}\\
     10\%  & \co{1872}{17.22}   & \co{3306}{67.83} & \co{5178}{83.27}   & \co{1954}{20.02}\\
     25\%  & \co{4692}{26.11}   & \co{5441}{61.11} & \co{10133}{83.45}  & \co{3858}{21.14}\\
     50\%  & \co{9396}{78.88}   & \co{5324}{67.09} & \co{14720}{143.22} & \co{5684}{36.78}\\
     75\%  & \co{14063}{109.61} & \co{3151}{36.92} & \co{17214}{144.08} & \co{6715}{30.44}\\
     100\% & \co{18806}{157.31} & 0 &              & \co{18806}{157.31} & \co{7367}{23.93}\\
    \end{tabular}
  \end{center}
  \caption{Random Selection}
  \label{sc.simu.res.rs}
\end{table*}

\begin{table*}[!t]
  \begin{center}
    \begin{tabular}{c|rrrrrrrr}
    & \multicolumn{2}{c}{$|B|$} & \multicolumn{2}{c}{$|B'|$} & \multicolumn{2}{c}{$|B|+|B'|$} & \multicolumn{2}{c}{$|A'|$}\\
      \hline
     1\%   & \co{188}{1.09}     & \co{431}{15.27}  & \co{619}{16.11}    & \co{183}{1.82}\\
     2\%   & \co{377}{2.75}     & \co{854}{18.14}  & \co{1231}{19.39}   & \co{362}{3.77}\\
     5\%   & \co{939}{7.67}     & \co{1942}{28.77} & \co{2881}{33.57}   & \co{857}{9.82}\\
     10\%  & \co{1877}{12.79}   & \co{3303}{65.26} & \co{5180}{76.46}   & \co{1577}{14.92}\\
     25\%  & \co{4667}{35.36}   & \co{5338}{72.65} & \co{10045}{105.28} & \co{3143}{19.83}\\
     50\%  & \co{9365}{44.51}   & \co{5330}{52.09} & \co{14695}{92.01}  & \co{4754}{24.27}\\
     75\%  & \co{14039}{85.94}  & \co{3128}{37.98} & \co{17167}{119.53} & \co{5710}{21.64}\\
     100\% & \co{18705}{173.76} & 0 &              & \co{18705}{173.76} & \co{6407}{36.02}\\
    \end{tabular}
  \end{center}
  \caption{Minimum FN Selection}
  \label{sc.simu.res.mfn}
\end{table*}

\begin{table*}[!t]
  \begin{center}
    \begin{tabular}{c|rrrrrrrr}
    & \multicolumn{2}{c}{$|B|$} & \multicolumn{2}{c}{$|B'|$} & \multicolumn{2}{c}{$|B|+|B'|$} & \multicolumn{2}{c}{$|A'|$}\\
      \hline
     1\%   & \co{187}{0.93}     & \co{769}{9.97}   & \co{956}{10.28}    & \co{226}{5.11}\\
     2\%   & \co{375}{1.82}     & \co{1458}{19.33} & \co{1833}{20.05}   & \co{447}{8.96}\\
     5\%   & \co{935}{6.36}     & \co{3154}{52.89} & \co{4089}{58.78}   & \co{1025}{12.08}\\
     10\%  & \co{1882}{16.55}   & \co{5188}{74.87} & \co{7070}{89.71}   & \co{1838}{20.53}\\
     25\%  & \co{4697}{34.52}   & \co{7466}{85.07} & \co{12163}{114.96} & \co{3420}{28.49}\\
     50\%  & \co{9396}{86.71}   & \co{6605}{98.04} & \co{16001}{182.14} & \co{4870}{29.84}\\
     75\%  & \co{14032}{99.42}  & \co{3670}{28.61} & \co{17702}{125.24} & \co{5674}{26.34}\\
     100\% & \co{18664}{138.13} & 0 &              & \co{18664}{138.13} & \co{6202}{22.09}\\
    \end{tabular}
  \end{center}
  \caption{Maximum FP Selection}
  \label{sc.simu.res.mfp}
\end{table*}

\begin{table*}[!t]
  \begin{center}
    \begin{tabular}{c|rrrrrrrr}
    & \multicolumn{2}{c}{$|B|$} & \multicolumn{2}{c}{$|B'|$} & \multicolumn{2}{c}{$|B|+|B'|$} & \multicolumn{2}{c}{$|A'|$}\\
      \hline
     1\%   & \co{188}{1.51}     & \co{735}{13.89}  & \co{923}{14.63}    & \co{188}{1.58}\\
     2\%   & \co{374}{3.25}     & \co{1372}{20.05} & \co{1746}{30.58}   & \co{363}{4.01}\\
     5\%   & \co{939}{6.92}     & \co{3035}{40.83} & \co{3974}{45.43}   & \co{844}{5.73}\\
     10\%  & \co{1863}{13.95}   & \co{4860}{67.65} & \co{6723}{78.71}   & \co{1498}{13.71}\\
     25\%  & \co{4703}{28.72}   & \co{7261}{68.39} & \co{11964}{94.59}  & \co{2895}{15.99}\\
     50\%  & \co{9394}{80.17}   & \co{6444}{70.86} & \co{15838}{149.01} & \co{4229}{25.95}\\
     75\%  & \co{14057}{126.61} & \co{3625}{38.28} & \co{17682}{162.64} & \co{5021}{27.54}\\
     100\% & \co{18683}{151.08} & 0 &              & \co{18683}{151.08} & \co{5581}{24.08}\\
    \end{tabular}
  \end{center}
  \caption{Ratio Selection}
  \label{sc.simu.res.ratio}
\end{table*}

Table~\ref{sc.simu.res.rs} to~\ref{sc.simu.res.ratio} present performance 
results for the selective clearing algorithms proposed in Sec.~\ref{sc.algo}.  
The mean over the fifteen run and the confidence intervals are shown.  The 
column $|B|$ gives the number of troublesome keys to remove.  The column $|B'|$ 
gives an idea of the side effect of performing selective clearing, in terms of 
additional false positive keys removed.  The column $|B + B'|$ shows the total 
number of false positive removed.  Finally, the last column, $|A'|$, 
illustrates the quantity of keys that become false negatives after selective 
clearing.

Looking first at the side effects (i.e., column $|B'|$), we see that removing
troublesome keys in $B$ has the consequence of removing other false positives.
Maximum FP Selection (Table~\ref{sc.simu.res.mfn}) and Ratio Selection
(Table~\ref{sc.simu.res.ratio}) have a larger side effect compared to the two
other selective clearing algorithms.  We further note that the total amount of
false positives removed from the filter (column $|B + B'|$) is larger than the
quantity of false negative generated (column $|A'|$).  This was expected, as
explained in Sec.~\ref{rbf.random}.

Looking now at the quantity of false negative generated, one can see that 
Minimum FN Selection (Table~\ref{sc.simu.res.mfn}) and Ratio Selection 
generates fewer false negatives than Maximum FP Selection and Random Selection.

Consequently, from these preliminary results, one concludes that the Ratio 
Selection algorithm provides better performance.  In the rest of this section, 
we will see if this conclusion is still valid when comparing the four selective 
algorithms in terms of the number of reset bits required to remove troublesome 
keys in $B$ and in terms of the $\chi$ metric.

\begin{figure}[t]
  \begin{center}
    \includegraphics[width=6.5cm]{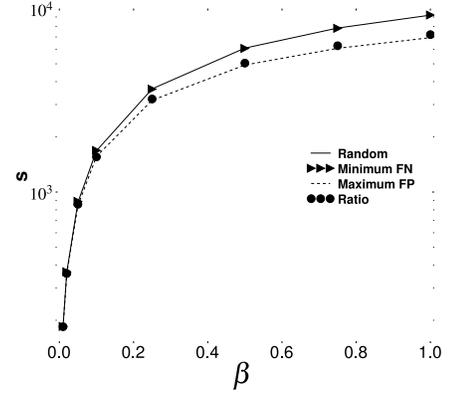}
  \end{center}
  \caption{Number of bits reset}
  \label{sc.simu.res.s}
\end{figure}

Fig.~\ref{sc.simu.res.s} compares the four algorithms in terms of the number 
$s$ of reset bits required to remove troublesome keys in $B$. The horizontal 
axis gives $\beta$ and the vertical axis, in log scale, gives $s$. The 
confidence intervals are plotted but they are too tight to appear clearly.

We see that Random Selection and Minimum FN Selection need to work more, in 
terms of number of bits to reset, when $\beta$ grows, compared to Maximum FP
Selection and Ratio Selection. In addition, we note that the Ratio Selection
algorithm needs to reset somewhat more bits than Maximum FP Selection (the
difference is too tight to be clearly visible on the plots).

\begin{figure}[!t]
  \begin{center}
    \includegraphics[width=6.5cm]{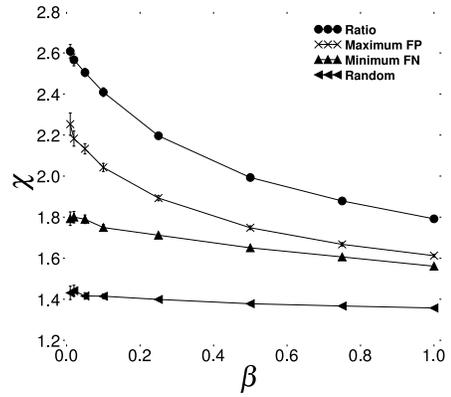}
  \end{center}
  \caption{Effect on $\chi$}
  \label{sc.simu.res.chi}
\end{figure}

Fig.~\ref{sc.simu.res.chi} evaluates the performance of the four algorithms. It
plots $\beta$ on the horizontal axis and $\chi$ on the vertical axis.  Again,
the confidence intervals are plotted but they are generally too tight to be
visible.

We first note that, whatever the algorithm considered, the $\chi$ ratio is 
always above 1, meaning that the advantages of removing false positives 
overcome the drawbacks of generating false negatives, if these errors are 
considered equally grave.  Thus, as expected, performing selective clearing 
provides better results than randomized bit clearing. Ratio Selection does 
best, followed by Maximum FP, Minimum FN, and Ratio Selection.

The $\chi$ ratio for Random Selection does not vary much with $\beta$ compared
to the three other algorithms.  For instance, the $\chi$ ratio for Ratio
Selection is decreased by 31.3\% between $\beta$=1\% and $\beta$=100\%.

To summarize, one can say that, when using RBF, one can reliably get a $\chi$
above 1.4, even when using a simple selective clearing algorithm, such as Random
Selection. Applying a more efficient algorithm, such as Ratio Selection, allows
one to get a $\chi$ above 1.8.  Such $\chi$ values mean that the proportion of
false positives removed is higher than the proportion of false negatives
generated.

In this section, we provided and evaluated four simple selective algorithms.  We
showed that two algorithms, Maximum FP Selection and Ratio Selection, are more
efficient in terms of number of bits to clear in the filter.  Among these two
algorithms, we saw that Ratio Selection provides better results, in terms of the
$\chi$ ratio.

\section{Improving Selective Clearing}\label{isc}
Sec.~\ref{sc} discussed four selective clearing algorithms.  Most of these
algorithms simplifies the selective clearing process by not updating the counting
vectors when a particular troublesome key is removed from the bit vector.   This
leads to an over-estimation of the quantity of false negatives generated at each
step, as well as a sub-estimation of the amount of false positives removed at
each step.

This section investigates improved selective clearing algorithms that keep up to
date the quantity of false negatives removed and false positives removed at each
step of the algorithms.

Sec.~\ref{isc.algo} discusses three improved selective algorithms;
Sec.~\ref{isc.theor} proposes a theorical analysis of the improved selective
algorithms; finally, Sec.~\ref{isc.simu} compares the performances of the
improved selective algorithms with the standard algorithms introduced in
Sec.~\ref{sc}.

\subsection{Algorithms}\label{isc.algo}
\begin{figure}[!t]
  \begin{center}
    \includegraphics[width=4.5cm]{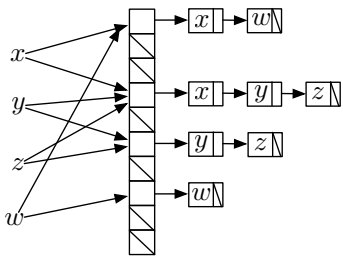}
  \end{center}
  \caption{Example of an ElementList vector}
  \label{isc.algo.elv}
\end{figure}

Our improved selective clearing algorithms, instead of using counting vectors, 
make use of a particular data structure illustrated in Fig.~\ref{isc.algo.elv}. 
We call such a data structure \dfn{ElementList vector}.  This is somewhat similar 
to the fast hash tables developed by Song et al.~\cite{extendedBF}.

The vector has the same length than the bit vector.  It contains thus $m$ cells.
Each cell is a pointer to a list of elements recorded in that position in the
bit vector.  These elements, depending on the selective clearing algorithm, can
belong to $A$ or $B$.

The first algorithm is an improvement to the Minimum FN Selection algorithm, 
called \dfn{Improved Minimum FN Selection}. Recall that Minimum FN Selection 
aims, for each troublesome key to remove, at selecting a bit amongst the $k$ 
available that will generate the minimum number of false negatives.  In the 
fashion of Minimum FN Selection, the minimum is given by the \textsc{MinIndex} 
procedure in Algorithm~\ref{isc.algo.mfn}. Instead of maintaining locally a 
counting vector, as done with the standard Minimum FN Selection algorithm, an 
ElementList vector, $v_A$, as illustrated in Fig.~\ref{isc.algo.elv}, is now used.  
Each cell of $v_A$ contains the list of elements belonging to $A$ that are 
recorded in the corresponding cell of $v$, the bit vector.  When the minimum 
index has been returned by \textsc{MinIndex}, the Improved Minimum FN Selection 
algorithm call the \textsc{BitClearing} procedure that will remove from $v_A$ 
all the elements recorded in this minimum index.  This was introduced in order 
to tackle the over-estimation of the standard Minimum FN Selection where the 
counting vector was not entirely updated at each step of the algorithm. 
Improved Minimum FN Selection is formally defined in 
Algorithm~\ref{isc.algo.mfn}.

\begin{algorithm}[!t]
  \caption{Improved Minimum FN Selection}
  \label{isc.algo.mfn}
  \begin{algorithmic}[1]
    \Require $v$, the bit vector and $v_A$, the element vector.
    \Ensure $v$ and $v_A$ updated, if needed.
    \Procedure{MinimumFNSelection}{$B$}
      \State \textsc{CreateCV}($A$)
      \ForAll{$b_i \in B$}
        \If{\textsc{MembershipTest}($b_i$, $v$)}
          \State index $\leftarrow$ \textsc{MinIndex}($b_i$)
          \State \textsc{BitClearing}($v_A$, index)
          \State $v$[index] $\leftarrow$ 0
        \EndIf
      \EndFor
    \EndProcedure
    \State
    \Procedure{CreateCV}{$A$}
      \ForAll{$a_i \in A$}
        \For{$j=1 \mathrm{\ to\ } k$}
          \State $v_A$[$h_j(a_i)$].add($a_i$)
        \EndFor
      \EndFor
    \EndProcedure
    \State
    \Procedure{BitClearing}{$\nu$, index}
      \State ElementList el = $\nu$.get(index)
      \ForAll{$x_i \in el$}
        \State \textsc{Remove}($x_i$, $\nu$)\Comment{remove all occurrences of
        element $x_i$ from $\nu$}
      \EndFor
    \EndProcedure
  \end{algorithmic}
\end{algorithm}

Note that the \textsc{MembershipTest} procedure is identical to the one
introduced in Sec.~\ref{sc.algo}.

The second improved selective clearing algorithm is an improvement to the 
Maximum FP Selection algorithm and is called \dfn{Improved Maximum FP 
Selection}.  The standard Maximum FP Selection algorithm, defined in 
Algorithm~\ref{sc.algo.mfp}, aims at removing the maximum quantity of 
troublesome false positives at each step of the algorithm.  Improved Maximum FP 
Selection behaves mainly in the same way, except it makes use of an element 
vector, $v_B$, instead of a counting vector. When the maximum index is found by 
the \textsc{MaxIndex} procedure, the \textsc{BitClearing} procedure is called in
order to maintain $v_B$ up-to-date. Improved Maximum FP Selection is formally
defined in Algorithm~\ref{isc.algo.mfp}.

\begin{algorithm}[!t]
  \caption{Improved Maximum FP Selection}
  \label{isc.algo.mfp}
  \begin{algorithmic}[1]
    \Require $v$, the bit vector and $v_B$, the element vector.
    \Ensure $v$ and $v_\textrm{B}$ updated, if needed.
    \Procedure{MaximumFP}{$B$}
      \State \textsc{CreateFV}($B$)
      \ForAll{$b_i \in B$}
        \If{\textsc{MembershipTest}($b_i$, $v$)}
          \State index $\leftarrow$ \textsc{MaxIndex}($b_i$)
          \State \textsc{BitClearing}($v_B$, index)
          \State $v$[index] $\leftarrow$ 0
        \EndIf
      \EndFor
    \EndProcedure
    \State
    \Procedure{CreateFV}{$B$}
      \ForAll{$b_i \in B$}
        \For{$j=1 \mathrm{\ to\ } k$}
          \State $v_B$[$h_j(b_i)$].add($b_i$)
        \EndFor
      \EndFor
    \EndProcedure
  \end{algorithmic}
\end{algorithm}

Finally, our last improved selective algorithms, the \dfn{Improved Ratio 
Selection} algorithm aims at increasing the performances of the standard Ratio 
Selection algorithm defined in Algorithm~\ref{sc.algo.ratio}.  Ratio Selection 
combines Minimum FN Selection and Maximum FP Selection into a single algorithm. 
It makes an attempt to minimize the false negatives generated while maximizing 
the false positives removed.  Improved Ratio Selection, in the spirit of our 
improved selective clearing algorithms, behaves the same way as its standard 
counterpart but it uses two ElementList vectors: $v_A$ that stores the elements 
belonging to $A$ and $v_B$ that stores the false positives recorded in $v$. 
These two ElementList vectors are maintained up-to-date thanks to the 
\textsc{BitClearing} procedure.  Further, the ratio vector, $r$, containing the 
ratio of the number of elements recorded in a given cell of $v_A$ to the number 
of elements recorded in a given cell of $v_B$ is also maintained up-to-date.  This
is achieved by calling the \textsc{Ratio} procedure each time a false positive
is removed from the bit vector.  The Improved Ratio Algorithm is formally
defined in Sec.~\ref{isc.algo.ratio}.

\begin{algorithm}[!t]
  \caption{Improved Ratio Selection}
  \label{isc.algo.ratio}
  \begin{algorithmic}[1]
    \Require $v$, the bit vector, $v_B$ and $v_A$, the ElementList
      vectors and $r$, the ratio vector.
    \Ensure $v$, $v_A$, $v_B$ and $r$ updated, if needed.
    \Procedure{Ratio}{$B$}
      \State \textsc{CreateCV}($A$)
      \State \textsc{CreateFV}($B$)
      \State \textsc{ComputeRatio}()
      \ForAll{$b_i \in B$}
        \If{\textsc{MembershipTest}($b_i$, $v$)}
          \State index $\leftarrow$ \textsc{MinRatio}($b_i$)
          \State \textsc{BitClearing}($v_A$, index)
          \State \textsc{BitClearing}($v_B$, index)
          \State $v$[index] $\leftarrow$ 0
          \State $r$[index] $\leftarrow$ 0
          \State \textsc{ComputeRatio}()
        \EndIf
      \EndFor
    \EndProcedure
    \State
    \Procedure{ComputeRatio}{}
      \For{$i=1 \mathrm{\ to\ } m$}
        \If{$v$[i] $\wedge$ $v_B[i].size() > 0$}
          \State $r$[i] $\leftarrow$ $\frac{v_A[i].size()}{v_B[i].size()}$
        \EndIf
      \EndFor
    \EndProcedure
  \end{algorithmic}
\end{algorithm}

\subsection{Theorical Analysis}\label{isc.theor}
\subsubsection{Algorithmic Analysis}\label{isc.theor.algo}
\begin{lem}
The running time of the Improved Minimum FN Selection algorithm is identical to 
the running time of the standard Minimum FN Selection algorithm, i.e.,  $O(k 
\times (|A| + |B|))$.
\end{lem}
\begin{proof}
Improved Minimum FN Selection starts by calling the \textsc{CreateCV} 
procedure. This procedure aims at creating the ElementList vector, $v_A$.  To 
do so, it browses all elements belonging to $A$ and each element is added $k$ 
times $v_A$.  As adding a cell to a list is an atomic operation (i.e., 
complexity $O(1)$), the algorithmic complexity of \textsc{CreateCV} is  $O(k 
\times |A|)$.

Improved Minimum FN Selection next browses all elements belonging to $B$ and, 
for each element, it performs the membership test.  If \textsc{MembershipTest} 
returns ``true'', then \textsc{MinIndex} (complexity $O(k)$, as demonstrated in 
Sec.~\ref{sc.theor.alg}) is called as well as \textsc{BitClearing}. Note that 
the algorithmic complexity of the cumulated calls of \textsc{BitClearing} 
cannot be worst than the algorithmic complexity of \textsc{CreateCV} (clearing 
the ElementList vector is not harder, in a complexity sense, than creating it).

Using the same reasoning than in Sec.~\ref{sc.theor.alg}, the algorithmic
complexity of Improved Minimum FN Selection is $O(\max(k \times |A|, 2k 
\times |B|))$, which leads to $O(k \times (|A| + |B|))$.
\end{proof}

\begin{lem}
The running time of the Improved Maximum FP Selection algorithm is identical to
the running time of the standard Maximum FP Selection algorithm, i.e., $O(k
\times |B|)$.
\end{lem}
\begin{proof}
Improved Maximum FP Selection starts by calling the \textsc{CreateFV} procedure
whose complexity is $O(k \times |B|)$.

Improved Maximum FP Selection next browses all elements belonging to $B$ and, 
for each element, it performs the membership test.  If \textsc{MembershipTest} 
returns ``true'', then \textsc{MaxIndex} (complexity $O(k)$, as demonstrated in 
Sec.~\ref{sc.theor.alg}) is called as well as \textsc{BitClearing}.  As stated
earlier in this section, the \textsc{BitClearing} complexity cannot be worst
than the ElementList vector creation. 

As a consequence, and using the same reasoning than in Sec.~\ref{sc.theor.alg},
the algorithmic complexity of Improved Maximum FP Selection is $O(k \times
|B|)$. 
\end{proof}

\begin{lem}
The running time of the Improved Ratio Selection algorithm is identical to the 
running time of the standard Ratio Selection algorithm, i.e., $O(k \times (|A| 
+ |B|) + m)$.
\end{lem}
\begin{proof}
After calling \textsc{CreateCV} (complexity $O(k \times |A|)$) and
\textsc{CreateFV} (complexity $O(k \times |B|)$), Improved Ratio Selection
called the \textsc{Ratio} procedure that aims at creating the ratio of $v_A$ to
$v_B$.  The complexity of \textsc{Ratio} is $O(m \times \max(|A|, |B|))$ as it must
browse all vector cells and, for each cell, count the number of elements
recorded in the list.

The rest of Improved Ratio Selection behaves the same way as Improved Minimum FN
Selection and Improved Maximum FP Selection, i.e., it browses all elements
belonging to $B$, performs the membership test and, if needed, selects the
minimum value among $k$ available.  Next, it maintains up-to-date $v_A$ and $v_B$
by calling \textsc{BitClearing}.  A new ratio is then calculated.

As a consequence, using the same reasoning than earlier in this section, the
algorithmic complexity of Improved Ratio Selection is $O(k \times (|A| + |B|) +
m)$. 
\end{proof}

\subsubsection{Spatial Complexity}\label{isc.theor.spat}
\begin{lem}
The spatial complexity of the Improved Minimum FN Selection algorithm is $O(m + 
|B| + k \times u \times |A|)$.
\end{lem}
\begin{proof}
The Improved Minimum FN Selection algorithm makes use of three data structures: 
the bit vector $v$, the set of troublesome keys to remove $B$ and the 
ElementList vector, $v_A$.  $v$ contains $m$ bits and $v_A$ contains, at worst, 
$k$ times each element belonging to $A$.  We finally consider that $u$ defines 
the space needed to store an element of the universe $U$ (and, consequently, of 
$A$). As a consequence, the spatial complexity of the Improved Minimum FN Selection 
algorithm is $O(m + |B| + k \times u \times |A|)$.
\end{proof}

\begin{lem}
The spatial complexity of the Improved Maximum FP Selection algorithm is $O(m + 
k \times u \times|B|)$.
\end{lem}
\begin{proof}
The Improved Maximum FP Selection algorithm makes use of three data structures: 
the bit vector $v$, the set of troublesome keys to remove $B$ and the 
ElementList vector, $v_B$.  $v$ contains $m$ bits and $v_B$ contains, at worst, 
$k$ times each troublesome key belonging to $B$.  Again, $u$ gives the space 
needed to store an element belonging to $U$. Consequently, the spatial 
complexity of the Improved Maximum FP Selection algorithm is $O(m + k \times u 
\times |B|)$.
\end{proof}

\begin{lem}
The spatial complexity of the Improved Ratio Selection algorithm is $O(m + k
\times u \times (|A| + |B|)+ dm)$.
\end{lem}
\begin{proof}
The Improved Ratio Selection algorithm makes use of five data structures: the
bit vector $v$, the set of troublesome keys to remove $B$, two ElementList
vectors, $v_A$ and $v_B$, and the ratio vector, $r$.  $v$ contains m bits, $v_A$
contains, at worst, $k$ times each element belonging to $A$, $v_B$ contains, at worst,
$k$ times each troublesome key belonging to $B$ and $r$ is a vector of $m$
floats (we consider that $d$ indicates the number of bits needed to store a
float).  Consequently, the spatial complexity of the Improved Ratio Selection
algorithm is $O(m + k \times u \times (|A| + |B|)+ dm)$.
\end{proof}

\subsection{Simulation Analysis}\label{isc.simu}
We conducted our simulations using the methodology explained in
Sec.~\ref{sc.simu.methodo}.  

\begin{figure*}[!htbp]
  \begin{minipage}[l]{0.33\linewidth}
    \begin{center}
      \includegraphics[width=6.5cm]{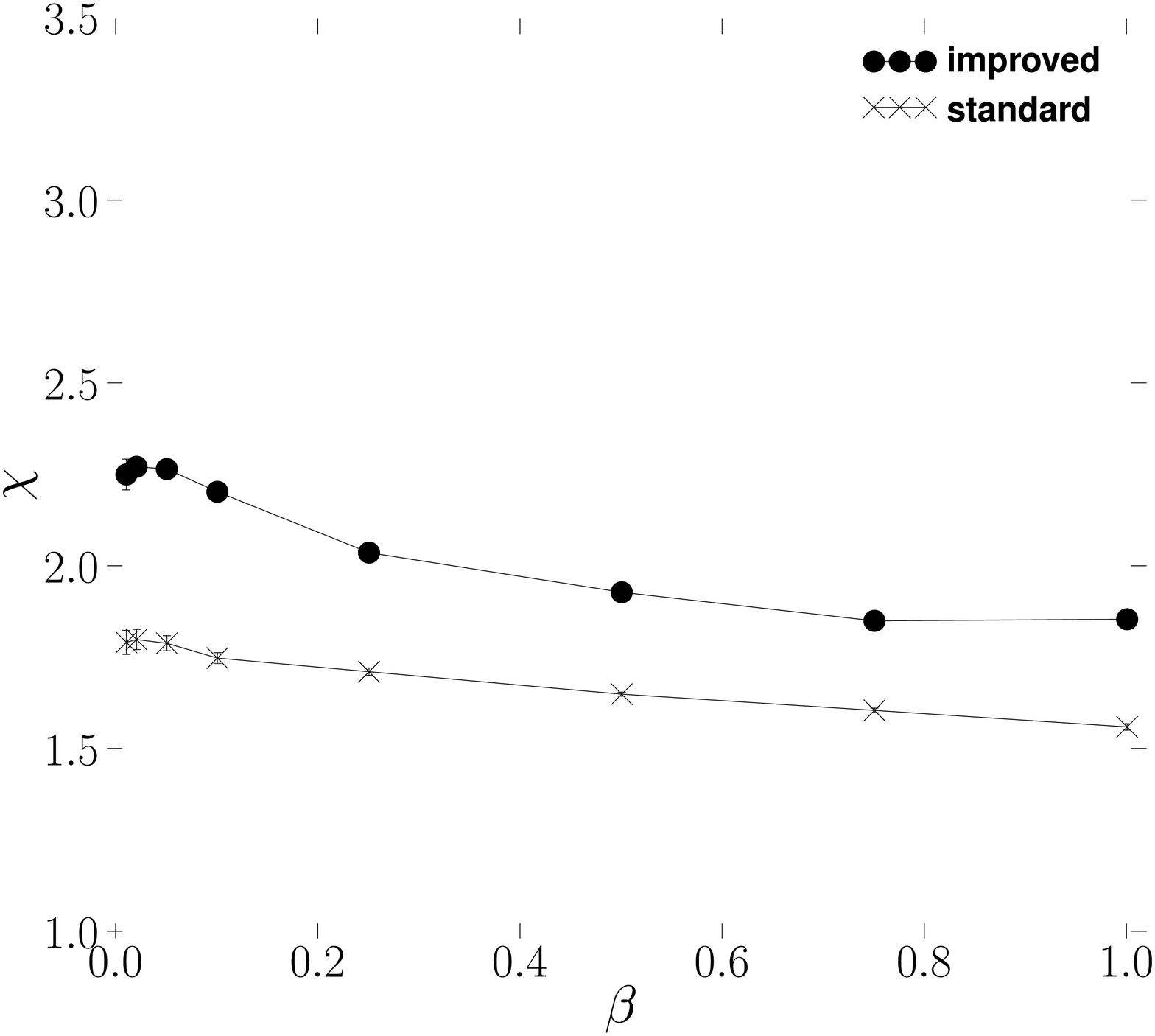}
      \caption{Minimum FN Selection comparison}
    \end{center}
    \label{isc.simu.min}
  \end{minipage}
  \begin{minipage}[l]{0.33\linewidth}
    \begin{center}
      \includegraphics[width=6.5cm]{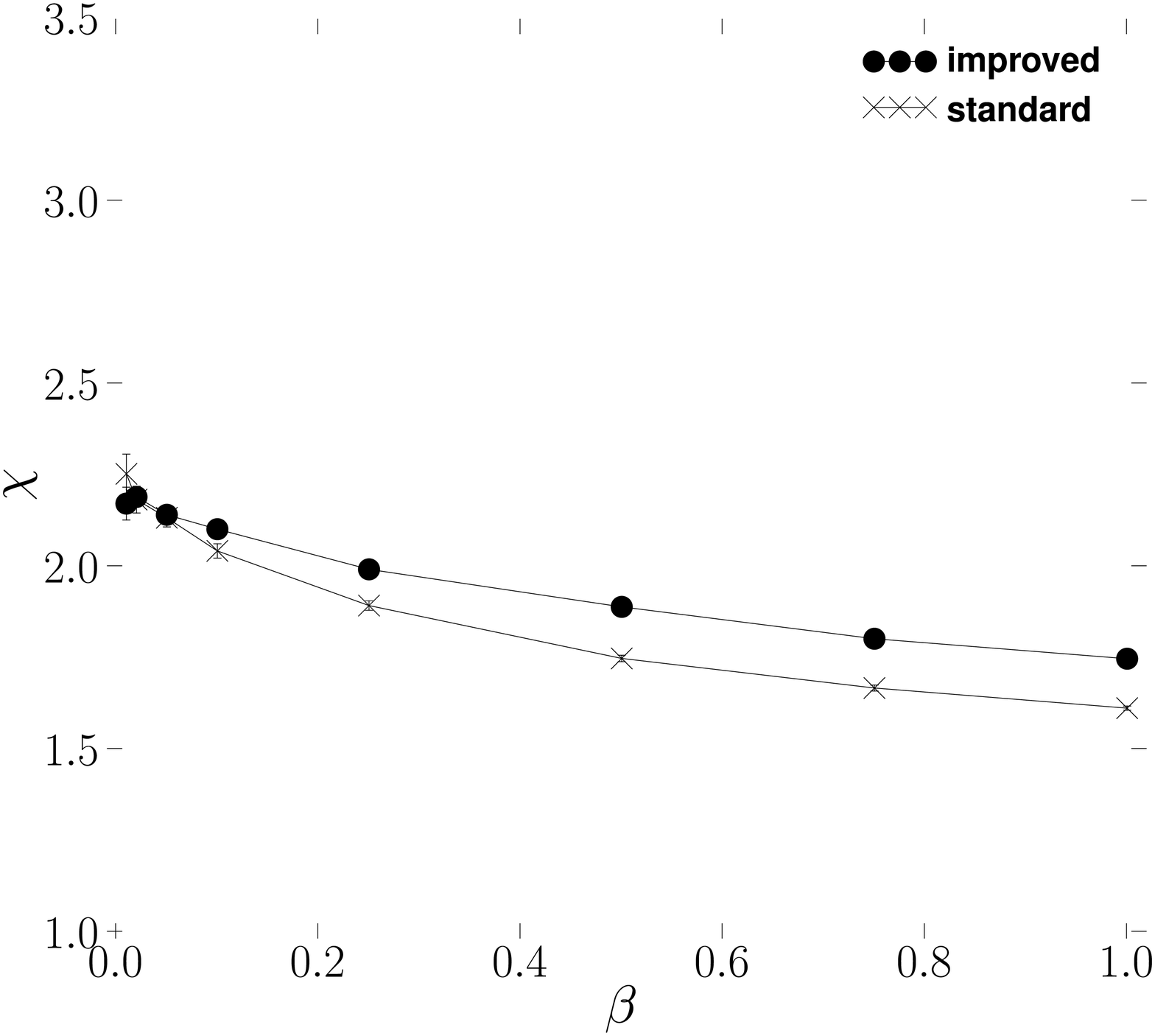}
      \caption{Maximum FP Selection comparison}
    \end{center}
    \label{isc.simu.max}
  \end{minipage}
  \begin{minipage}[l]{0.33\linewidth}
    \begin{center}
      \includegraphics[width=6.5cm]{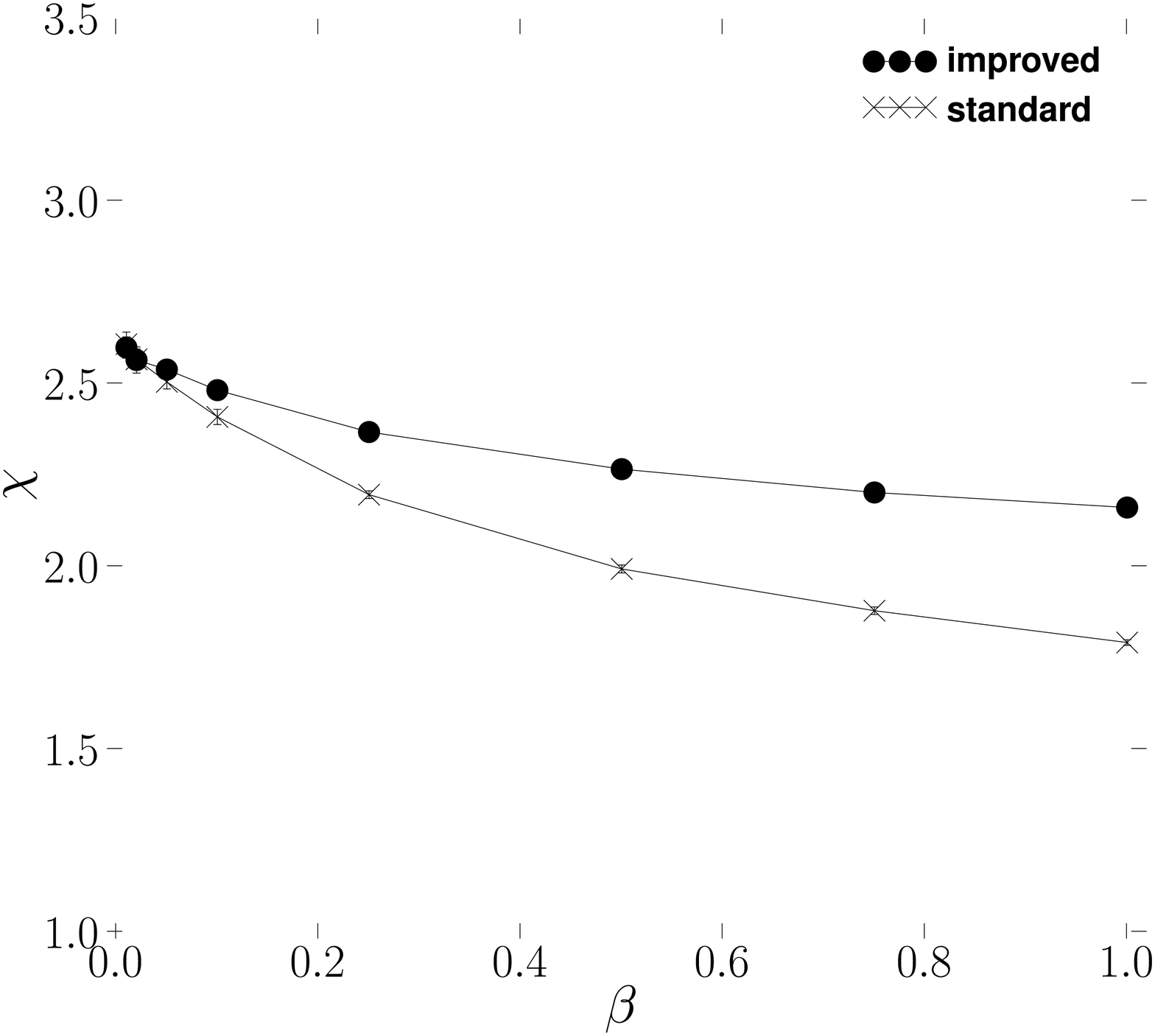}
      \caption{Ratio Selection comparison}
    \end{center}
    \label{isc.simu.rat}
  \end{minipage}
\end{figure*}

Fig.~\ref{isc.simu.min} to Fig.~\ref{isc.simu.rat} compare the 
performances of our improved selective clearing algorithms to the standard 
selective clearing algorithms.  The horizontal axis shows $\beta$, the ratio of 
the quantity of troublesome keys to remove to the whole false positive set (see 
eqn.~\ref{sc.eval.methodo.beta}).  The vertical axis gives $\chi$, the ratio 
between the proportion of false positives removed and the proportion of false 
negatives generated (see eqn.~\ref{bf.chi}).

We see that our improved selective clearing algorithms perform better than 
those described in Sec.~\ref{sc}.  In particular, Improved Minimum FN Selection 
provides the strongest increase compared to the standard algorithm:  between 
66.048\% ($\beta=0.01$) and 84.129\% ($\beta=0.75$).  Improved Maximum FN
Selection and Improved Ratio Selection provides better results, compared to the
standard version of the algorithms, when $\beta$ is high. Finally, Improved
Ratio Selection provides the best results, as expected from standard selective 
clearing algorithms.

\begin{figure}[!t]
  \begin{center}
    \subfigure[Filled proportion of the ElementList vector]{\label{isc.simu.el.vector}
      \includegraphics[width=6.5cm]{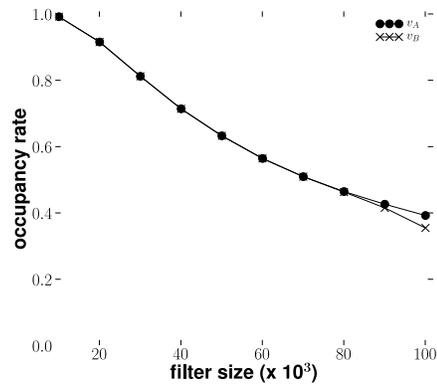}}
    \subfigure[Average ElementList size]{\label{isc.simu.el.list}
      \includegraphics[width=6.5cm]{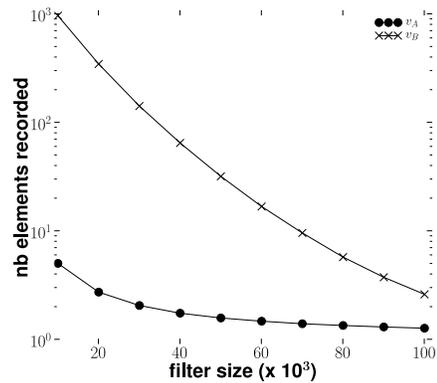}}
  \end{center}
  \caption{ElementList vector evaluation}
  \label{isc.simu.el}
\end{figure}

Fig.~\ref{isc.simu.el} evaluates the ElementList vector data structure.  The
horizontal axis, for both plots, gives the vector size, $m$.  We vary it between
10$^4$ and 10$^5$, with an increment of 10$^4$.

Fig.~\ref{isc.simu.el.vector} shows, on the vertical axis, the proportion of the
vector that is used. If it is equal to 1, it means that all cells in the vector
contain, at least, one element.  Otherwise, if it is equal to 0, it means that
all the vector cells are empty.  We see from Fig.~\ref{isc.simu.el.vector} that
the occupancy rate of the vector decreases nearly linearly with the vector size.
We also notice that the occupancy rate of both vector, $v_A$ and $v_B$, is the
same for most of the vector sizes.  When the vector is larger, i.e., above
90,000 cells, the occupancy rate of $v_B$ becomes smaller than $v_A$.

Fig.~\ref{isc.simu.el.list} shows, on the vertical axis, the average size of an
ElementList item in the vector.  The minimum size is 1 (otherwise, the list is empty).
The maximum value is either $|A|$, for $v_A$, either $|B|$, for $v_B$.  For our
experiments, we consider that $B$ equals $FP$.  Looking first at the $v_B$
vector, one can see that the average ElementList size decreases quickly when the
vector size increase.  It decreases by two order of magnitude while the vector
size increases only by one order.  Looking now at the $v_A$ vector, we see that
in the worst case, a cell contains, on average, less than ten elements.  It
quickly decreases until having, at worst, one element per filled cell.

\section{Case Study}\label{case}
\subsection{Tracing Paths with a Red Stop Set}\label{case.rss}
Retouched Bloom filters can be applied across a wide range of applications that 
would otherwise use Bloom filters.  For RBFs to be suitable for an application, 
two criteria must be satisfied.  First, the application must be capable of 
identifying instances of false positives.  Second, the application must accept 
the generation of false negatives, and in particular, the marginal benefit of 
removing the false positives must exceed the marginal cost of introducing the 
false negatives.

This section describes the application that motivated our introduction of RBFs: 
a network measurement system that traces routes, and must communicate 
information concerning IP addresses at which to stop tracing. 
Sec.~\ref{related.app} will investigate others applications that can benefit 
from RBFs instead of Bloom filters. Sec.~\ref{case.eval} evaluates the impact 
of using RBFs in this application.

Maps of the internet at the IP level are constructed by tracing routes from
measurement points distributed throughout the internet.  The \dfn{skitter}
system~\cite{skitter}, which has provided data for many network topology
papers, launches probes from 24 monitors towards almost a million destinations.
However, a more accurate picture can potentially be built by using a larger
number of vantage points. \textsc{Dimes}~\cite{dimes} heralds a new generation
of large-scale systems, counting, at present 8,700 agents distributed over five
continents.  As Donnet et al.~\cite{DonnetEfficient2005} (including authors on
the present paper) have pointed out, one of the dangers posed by a large number
of monitors probing towards a common set of destinations is that the traffic
may easily be mistaken for a distributed denial of service (DDoS) attack.

One way to avoid such a risk would be to avoid hitting destinations. This can
be done through smart route tracing algorithms, such as Donnet et al.'s
\dfn{Doubletree}.  With Doubletree, monitors communicate amongst themselves
regarding routes that they have already traced, in order to avoid duplicating 
work.  Since one monitor will stop tracing a route when it reaches a point that
another monitor has already traced, it will not continue through to hit the
destination.

\begin{figure}[!t]
  \begin{center}
    \includegraphics[width=5.5cm]{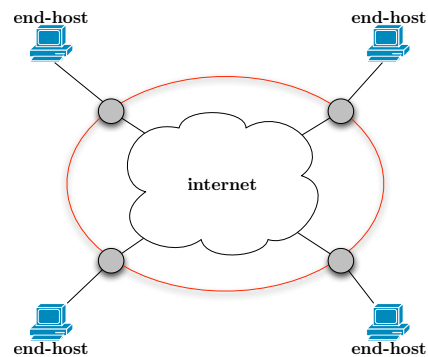}
  \end{center} 
  \caption{Red stop set}
  \label{case.rss.fig}
\end{figure}

Doubletree considerably reduces, but does not entirely eliminate, DDoS risk. 
Some monitors will continue to hit destinations, and will do so repeatedly. One
way to further scale back the impact on destinations would be to introduce an
additional stopping rule that requires any monitor to stop tracing when it 
reaches a node that is one hop before that destination.  We call such a node
the \dfn{penultimate node}, and we call the set of penultimate nodes the 
\dfn{red stop set} (RSS).Fig.~\ref{case.rss.fig} illustrates the RSS concept,
showing penultimate nodes as grey discs.

A monitor is typically not blocked by its own first-hop node, 
as it will normally see a different IP address from the addresses that appear 
as penultimate nodes on incoming traces.  This is because a router has multiple
interfaces, and the IP address that is revealed is supposed to be the one that
sends the probe reply.  The application that we study in this paper conducts
standard route traces with an RSS.  We do not use Doubletree, so as to avoid 
having to disentangle the effects of using two different stopping rules at the
same time.

How does one build the red stop set? The penultimate nodes cannot be determined
a priori. However, the RSS can be constructed during a learning round in which
each monitor performs a full set of standard traceroutes, i.e., until hitting a
destination. Monitors then share their RSSes.  For simplicity, we consider that
they all send their RSSes to a central server, which combines them to form a 
global RSS, that is then redispatched to the monitors.  The monitors then apply
the global RSS in a stopping rule over multiple rounds of probing.

Destinations are only hit during the learning round and as a result of errors
in the probing rounds.  DDoS risk diminishes with an increase in the ratio of
probing rounds to learning rounds, and with a decrease in errors during the
probing rounds.  DDoS risk would be further reduced were we to apply Doubletree
in the learning round, as the number of probes that reach destinations during
the learning round would then scale less then linearly in the number of
monitors.  However, our focus here is on the probing rounds, which use the
global RSS, and not on improving the efficiency of the learning round, which
generates the RSS, and for which we already have known techniques.

The communication cost for sharing the RSS among monitors is linear in the 
number of monitors and in the size of the RSS representation.  It is this 
latter size that we would like to reduce by a constant compression factor.  If 
the RSS is implemented as a list of 32-bit vectors, skitter's million 
destinations would consume 4 MB.  We therefore propose encoding the RSS 
information in Bloom filters.  Note that the central server can combine 
similarly constructed Bloom filters from multiple monitors, through bitwise 
logical \textsc{or} operations, to form the filter that encodes the global RSS.

The cost of using Bloom filters is that the application will encounter false
positives. A false positive, in our case study, corresponds to an early stop in
the probing, i.e., before the penultimate node. We call such an error
\dfn{stopping short}, and it means that part of the path that should have been
discovered will go unexplored.  Stopping short can also arise through network
dynamics, when additional nodes are introduced, by routing changes or IP
address reassignment, between the previously penultimate node and the
destination.  In contrast, a trace that stops at a penultimate node is deemed a
\dfn{success}.  A trace that hits a destination is called a \dfn{collision}.
Collisions might occur because of a false negative for the penultimate node, or
simply because routing dynamics have introduced a new path to the destination,
and the penultimate node on that path was previously unknown.

As we show in Sec.~\ref{case.eval}, the cost of stopping short is far from
negligible.  If a node that has a high betweenness centrality (Dall'Asta et
al.~\cite{dallAsta} point out the importance of this parameter for topology
exploration) generates a false positive, then the topology information loss
might be high.  Consequently, our idea is to encode the RSS in an RBF.

There are two criteria for being able to profitably employ RBFs, and they are
both met by this application.  First, false positives can be identified and
removed.  Once the topology has been revealed, each node can be tested against
the Bloom filter, and those that register positive but are not penultimate
nodes are false positives. The application has the possibility of removing the
most troublesome false positives by using one of the selective algorithms
discussed in Sec.~\ref{sc}.  Second, a low rate of false negatives is
acceptable and the marginal benefit of removing the most troublesome false
positives exceeds the marginal cost of introducing those false negatives.  Our
aim is not to eliminate collisions; if they are considerably reduced, the DDoS
risk has been diminished and the RSS application can be deemed a success.  On
the other hand, systematically stopping short at central nodes can severely
restrict topology exploration, and so we are willing to accept a low rate of
random collisions in order to trace more effectively.  These trade-offs are
explored in the Sec.~\ref{case.eval}.

\begin{table*}[!t]
  \begin{center}
    \begin{tabular}{l|cccc|ccc}
     \textbf{Implementation} & \multicolumn{4}{c}{\textbf{Positive}} & \multicolumn{3}{c}{\textbf{Negative}}\\
     & Success & Topo.~discovery & Compression & No Collision & Topo.~missed & Load & Collision\\
     \hline
     List         & X & X &   & X &   & X & \\
     Bloom filter &   &   & X & X & X &   & \\
     RBF          & X & X & X &   &   &   & X\\
    \end{tabular}
  \end{center}
  \caption{Positive and negative aspects of each RSS implementation}
  \label{case.rss.tab}
\end{table*}

Table~\ref{case.rss.tab} summarizes the positive and negative aspects of each
RSS implementation we propose.  Positive aspects are a success, stopping at
the majority of penultimate nodes, topology information discovered, the
eventual compression ratio of the implementation and a minimum number of
collisions with destinations.   Negative aspects of an implementation can be
the topology information missed due to stopping short, the load on the network
when exchanging the RSS and the risk of hitting destinations too much times.
Sec.~\ref{case.eval} will measure the positive and negative aspects of each
implementation.

\subsection{Evaluation}\label{case.eval}
In this section, we evaluate the use of RBFs in a tracerouting system based on an
RSS.  We first present our methodology and then, 
discuss our results.

\subsubsection{Methodology}\label{case.eval.methodo}
Our study was based on skitter data~\cite{skitter} from January 2006.  This data
set was generated by 24 monitors located in the United States of America,
Canada, the United Kingdom, France, Sweden, the Netherlands, Japan, and New
Zealand. The monitors share a common destination set of 971,080 IPv4 addresses.
Each monitor cycles through the destination set at its own rate, taking
typically three days to complete a cycle.

For the purpose of our study, in order to reduce computing time to a manageable
level, we worked from a limited set of 10 skitter monitors, all the monitors
sharing a list of 10,000 destinations, randomly chosen from the original set.
In our data set, the RSS contains 8,006 different IPv4 addresses.

We will compare the three RSS implementations discussed above: list, Bloom
filter and RBF.  The list would not return any errors if the network were
static, however, as discussed above, network dynamics lead to a certain error
rate of both collisions and instances of stopping short.

For the RBF implementation, we considered $\beta$ values (see eqn. 
\ref{sc.eval.methodo.beta}) of 1\%, 5\%, 10\% and 25\%.  We further applied the 
Ratio Selection algorithm, as defined in Sec.~\ref{sc.algo}. For the Bloom 
filter and RBF implementations, the hashing was emulated with random numbers. 
We simulate randomness with the Mersenne Twister MT19937 pseudo-random number 
generator~\cite{twister}.

To obtain our results, we simulated one learning round on a first cycle of 
traceroutes from each monitor, to generate the RSS.  We then simulated one 
probing round, using a second cycle of traceroutes.  In this simulation, we 
replayed the traceroutes, but applied the stopping rule based on the RSS, 
noting instances of stopping short, successes, and collisions.

\subsubsection{Results}\label{case.eval.res}
\begin{figure}[!t]
  \begin{center}
      \includegraphics[width=6.5cm]{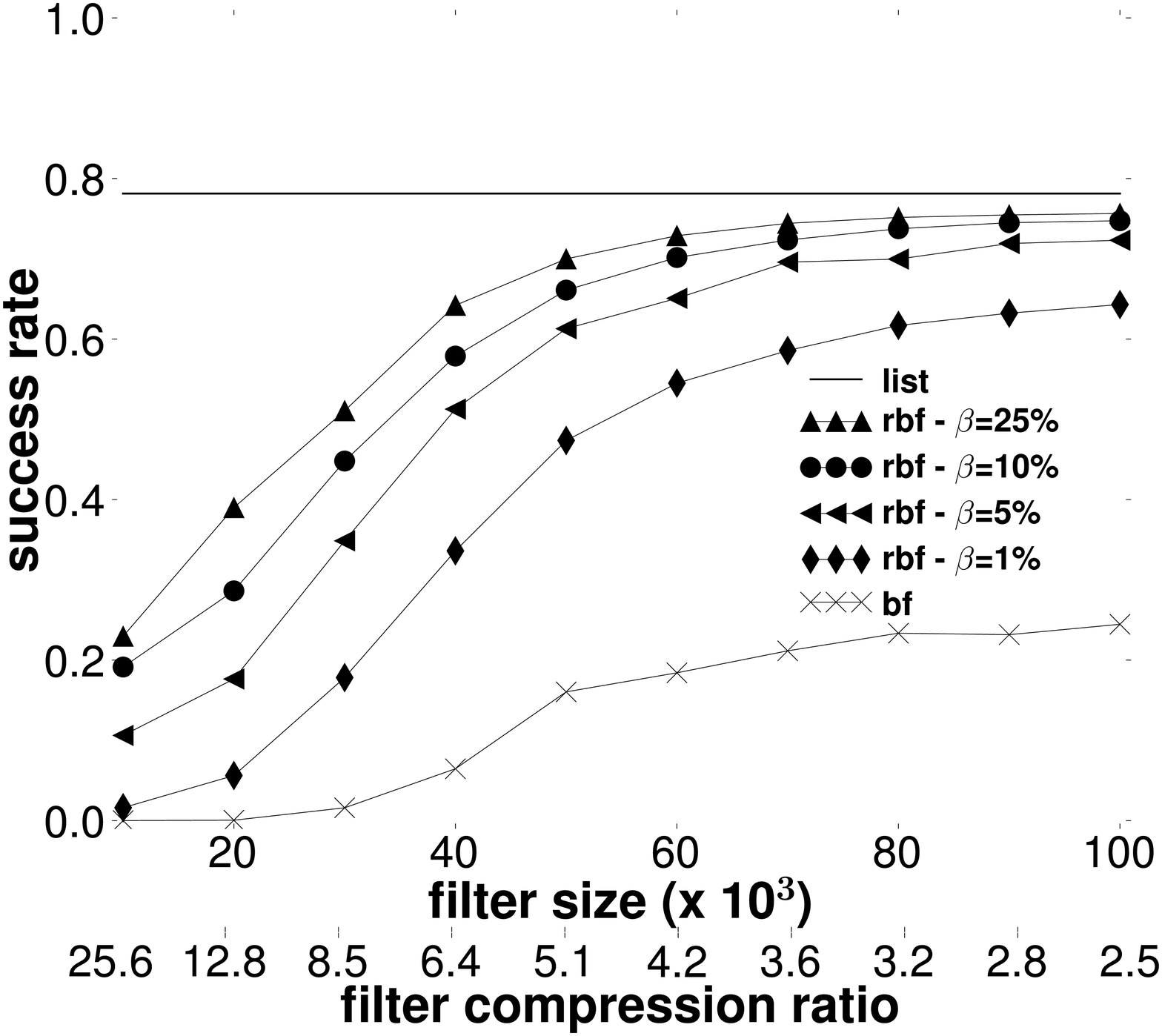}
  \end{center}
  \caption{Success rate}
  \label{case.eval.res.succ}
\end{figure} 

Fig.~\ref{case.eval.res.succ} compares the success rate, i.e., stopping at a
penultimate node, of the three RSS implementations.  The horizontal axis gives
different filters size, from 10,000 to 100,000, with an increment of 10,000. 
Below the horizontal axis sits another axis that indicates the compression
ratio of the filter, compared to the list implementation of the RSS. The 
vertical axis gives the success rate.  A value of 0 would mean that using a
particular implementation precludes stopping at the penultimate node. On the
other hand, a value of 1 means that the implementation succeeds in stopping
each time at the penultimate node.

Looking first at the list implementation (the horizontal line), we see that the
list implementation success rate is not 1 but, rather, 0.7812.  As explained in
Sec.~\ref{case.eval}, this can be explained by the network dynamics such
as routing changes and dynamic IP address allocation.

\begin{figure}[!t]
  \begin{center}
        \includegraphics[width=6.5cm]{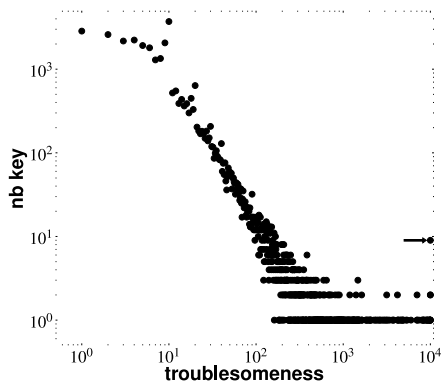}
  \end{center}
  \caption{Troublesomeness distribution}
  \label{case.eval.res.trouble} 
\end{figure}

With regards to the Bloom filter implementation, we see that the results are 
poor.  The maximum success rate, 0.2446, is obtained when the filter size is
100,000 (a compression ratio of 2.5 compared to the list).  Such poor results 
can be explained by the troublesomeness of false positives. 
Fig.~\ref{case.eval.res.trouble} shows, in log-log scale, the troublesomeness
distribution of false positives.  The horizontal axis gives the 
\dfn{troublesomeness degree}, defined as the number of traceroutes that stop
short for a given key. The maximum value is $10^4$, i.e., the number of 
traceroutes performed by a monitor.  The vertical axis gives the number of 
false positive elements having a specific troublesomeness degree.  The most 
troublesome keys are indicated by an arrow towards the lower right of the 
graph: nine false positives are, each one, encountered 10,000 times.
      
Looking now, in Fig.~\ref{case.eval.res.succ}, at the success rate of the RBF,
we see that the maximum success rate is reached when $\beta$ = 0.25. We also
note a significant increase in the success rate for RBF sizes from 10,000 to
60,000.  After that point, except for $\beta$ = 1\%, the increase is less 
marked and the success rate converges to the maximum, 0.7564.  When $\beta$ =
0.25, for compression ratios of 4.2 and lower, the success rate approaches 
that of the list implementation. Even for compression ratios as high as 25.6,
it is possible to have a success rate over a quarter of that offered by the
list implementation.

\begin{figure}[!t]
  \begin{center}
    \includegraphics[width=6.5cm]{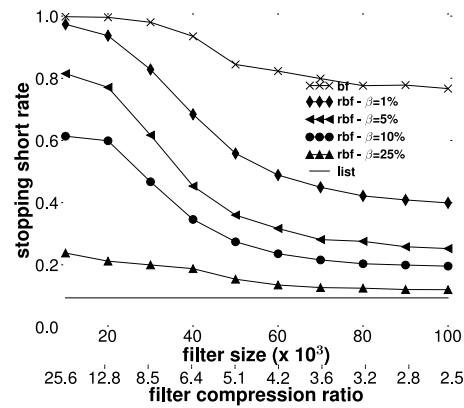}
  \end{center}
  \caption{Stopping short rate}
  \label{case.eval.res.ss}
\end{figure}

Fig.~\ref{case.eval.res.ss} gives the stopping short rate of the three RSS
implementations.  A value of 0 means that the RSS implementation does not 
generate any instances of stopping short.  On the other hand, a value of 1
means that every stop was short.

Looking first at the list implementation, one can see that the stopping short
rate is 0.0936.  Again, network dynamics imply that some nodes that were
considered as penultimate nodes during the learning phase are no longer
located one hop before a destination.

\begin{figure}[!t]
  \begin{center}
    \subfigure[nodes]{\label{case.eval.res.topo.nodes}
      \includegraphics[width=6.5cm]{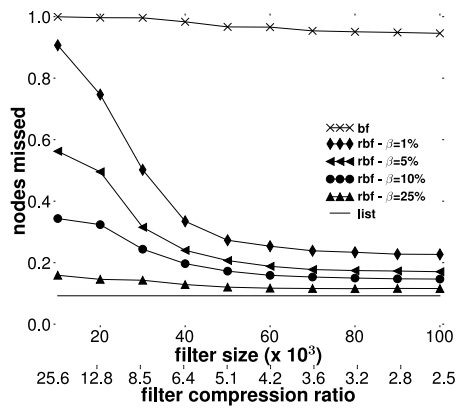}}
    \subfigure[links]{\label{case.eval.res.topo.links}
      \includegraphics[width=6.5cm]{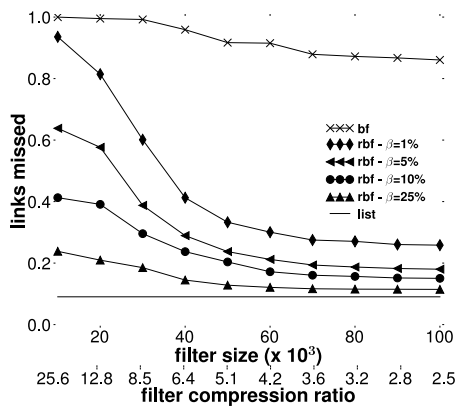}}
  \end{center}
  \caption{Topology information missed}
  \label{case.eval.res.topo}
\end{figure}

Regarding the Bloom filter implementation, one can see that the stopping short 
rate is significant.  Between 0.9981 (filter size of 10$^3$) and 0.7668 (filter
size of 10$^4$).  The cost of these high levels of stopping short can be
evaluated in terms of topology information missed. 
Fig.~\ref{case.eval.res.topo} compares the RBF and the Bloom filter 
implementation in terms of nodes (Fig.~\ref{case.eval.res.topo.nodes}) and 
links (Fig.~\ref{case.eval.res.topo.links}) missed due to stopping short. A 
value of 1 means that the filter implementation missed all nodes and links when
compared to the list implementation.  On the other hand, a value of 0 mean that
there is no loss, and all nodes and links discovered by the list implementation
are discovered by the filter implementation.  One can see that the loss, when
using a Bloom filter, is above 80\% for filter sizes below 70,000.

Implementing the RSS as an RBF allows one to decrease the stopping short rate.
When removing 25\% of the most troublesome false positives, one is able to
reduce the stopping short between 76.17\% (filter size of 10$^3$) and 84,35\%
(filter size of 10$^4$). Fig.~\ref{case.eval.res.ss} shows the advantage of
using an RBF instead of a Bloom filter. Fig.~\ref{case.eval.res.topo} shows
this advantage in terms of topology information.  We miss a much smaller
quantity of nodes and links with RBFs than Bloom filters and we are able to
nearly reach the same level of coverage as with the list implementation.

\begin{figure}[t]
  \begin{center}
    \includegraphics[width=6.5cm]{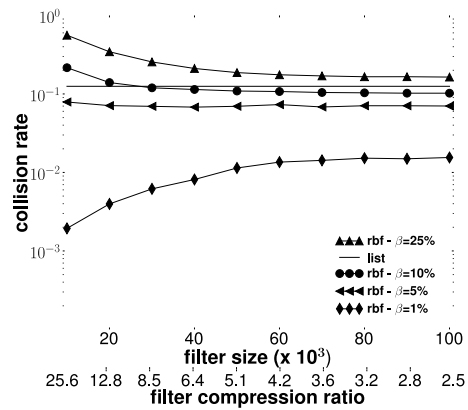}
  \end{center}
  \caption{Collision cost}
  \label{case.eval.res.cost}
\end{figure}

Fig.~\ref{case.eval.res.cost} shows the cost in terms of collisions.
Collisions will arise under Bloom filter and list implementations only due to
network dynamics.  Collisions can be reduced under all RSS implementations due
to a high rate of stopping short (though this is, of course, not desired).  The
effect of stopping short is most pronounced for RBFs when $\beta$ is low, as
shown by the curve $\beta = 0.01$.  One startling revelation of this figure is
that even for fairly high values of $\beta$, such as $\beta = 0.10$, the effect
of stopping short keeps the RBF collision cost lower than the collision cost
for the list implementation, over a wide range of compression ratios.  Even at
$\beta = 0.25$, the RBF collision cost is only slightly higher.

\begin{figure}[!t]
  \begin{center}
    \includegraphics[width=6.5cm]{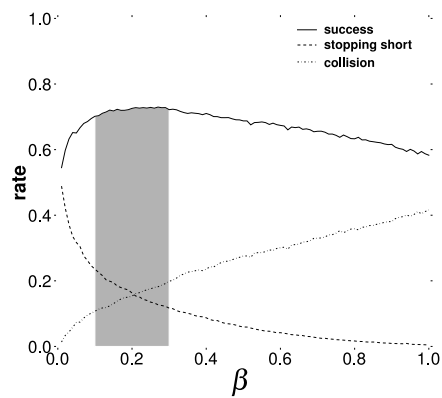}
  \end{center}
\vspace{-0.6cm}
  \caption{Metrics for an RBF with m=60,000}
  \label{case.eval.res.comparison}
\end{figure}

Fig.~\ref{case.eval.res.comparison} compares the success, stopping short, and
collision rates for the RBF implementation with a fixed filter size of 60,000
bits. We vary $\beta$ from 0.01 to 1 with an increment of 0.01.  We see that
the success rate increases with $\beta$ until reaching a peak at 0.642 ($\beta$
= 0.24), after which it decreases until the minimum success rate, 0.4575, is
reached at $\beta$ = 1. As expected, the stopping short rate decreases with
$\beta$, varying from 0.6842 ($\beta$ = 0) to 0 ($\beta$ = 1).  On the other
hand, the collision rate increases with $\beta$, varying from 0.0081 ($\beta$ =
0) to 0.5387 ($\beta$ = 1).

The shaded area in Fig.~\ref{case.eval.res.comparison} delimits a range of
$\beta$ values for which success rates are highest, and collision rates are
relatively low.  This implementation gives a compression ratio of 4.2 compared
to the list implementation.  The range of $\beta$ values (between 0.1 and 0.3)
gives a success rate between 0.7015 and 0.7218 while the list provides a
success rate of 0.7812.  The collision rate is between 0.1073 and 0.1987,
meaning that in less than 20\% of the cases a probe will hit a destination.  On
the other hand, a probe hits a destination in 12.51\% of the cases with the
list implementation. Finally, the stopping short rate is between 0.2355 and
0.1168 while the list implementation gives a stopping short rate of 0.0936.

\begin{figure}[!t]
  \begin{center}
    \includegraphics[width=6.5cm]{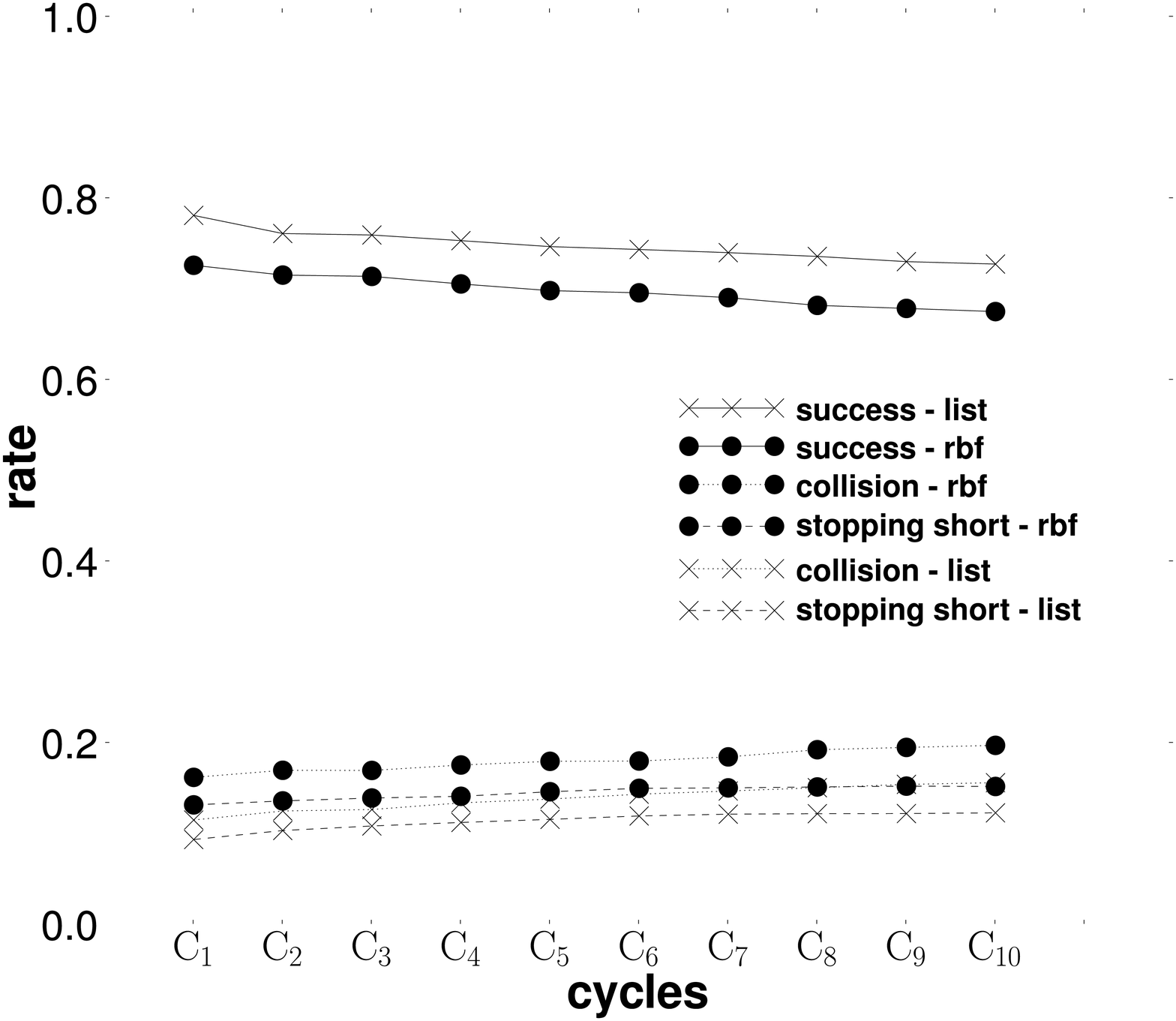}
  \end{center}
  \caption{Metrics for 10 traceroute cycles}
  \label{case.eval.res.cycles}
\end{figure}

Fig.~\ref{case.eval.res.cycles} illustrates the behavior of the RSS
during ten traceroute cycle.  We consider the list and the RBF implementations.
The RBF is tuned as followed: the vector is 60,000 bits long and $\beta$ is
0.25.  These values are suggested by previous studies in this section.  The
horizontal axis, in Fig.~\ref{case.eval.res.cycles}, gives the ten cycles, the
cycle labeled $C_1$ is equivalent to the results discussed below. The
vertical axis gives the metric rate (i.e., success, stopping short and
collision). 

In Fig.~\ref{case.eval.res.cycles}, one can see the degradation of the RSS
performances.  The success rate decreases with time while the stopping short and
the collision rates increases with time. However, both implementation behaves in
the same way.  The decrease of the success rate for the RBF is somewhat similar
to the list one.  The same conclusion holds for the stopping short and collision
rates.  Fig.~\ref{case.eval.res.cycles} shows thus the robustness of the RBF.

In closing, we emphasize that the construction of $B$ and the choice of $\beta$ 
in this case study are application specific.  We do not provide guidelines for 
a universal means of determining which false positives should be considered 
particularly troublesome, and thus subject to removal, across all applications. 
However, it should be possible for other applications to measure, in a similar 
manner as was done here, the potential benefits of introducing RBFs.

\subsection{Comparing Selective Clearing Algorithms}\label{case.compare}
In this section, we compare the performances of the Ratio Selection algorithm and
the Improved Ratio Selection algorithm for our case study.  The methodology
applied was the same than the one described in Sec.~\ref{case.eval.methodo},
except that we did not consider the Bloom filter implementation of the RSS.  In
order to make the plots readable, we only took into account $\beta=0.01$ and
$\beta=0.25$. 

\begin{figure}[!t]
  \begin{center}
    \includegraphics[width=6.5cm]{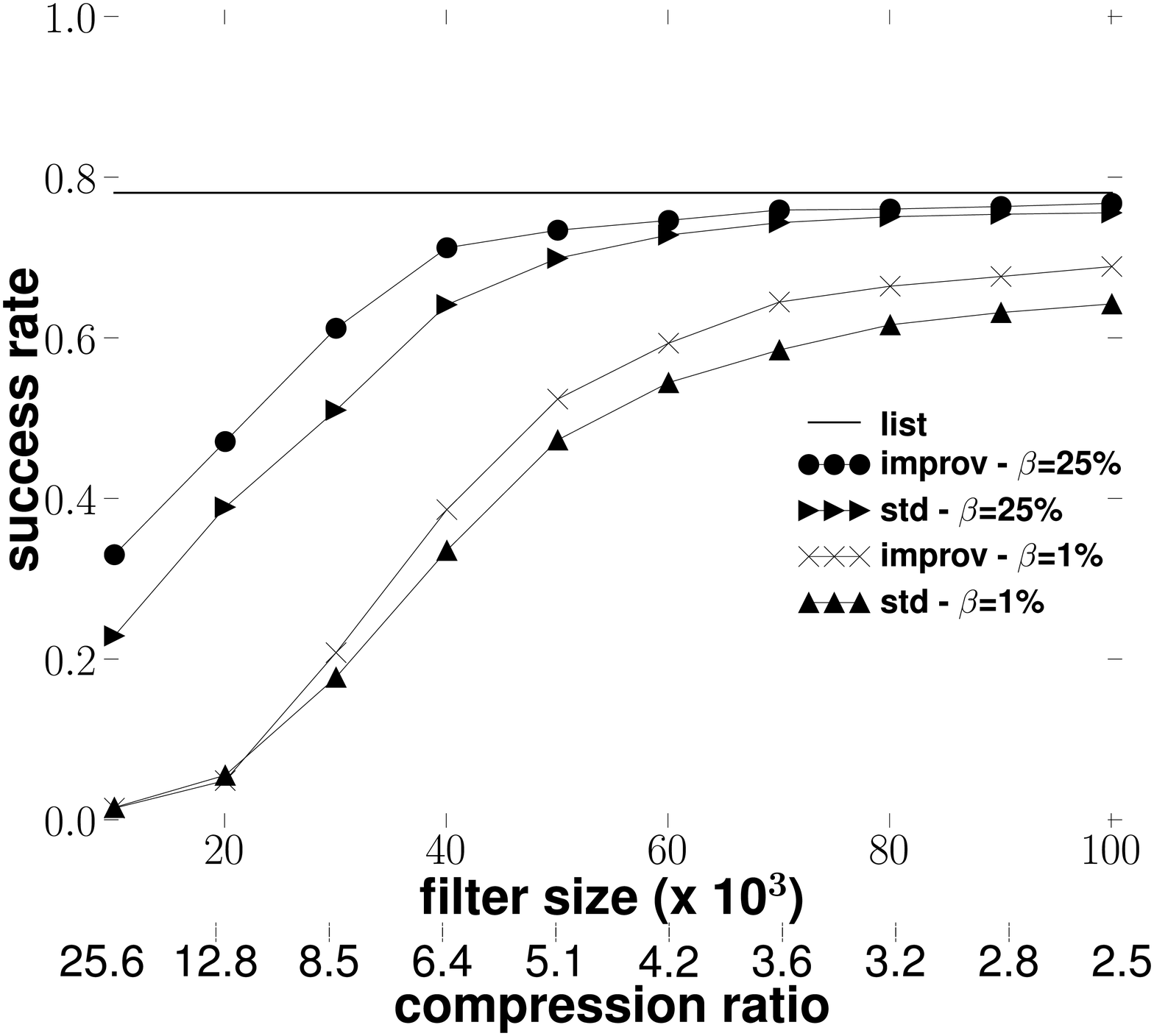}
  \end{center}
  \caption{Success comparison}
  \label{case.compare.success}
\end{figure}

Fig.~\ref{case.compare.success} compares both selective clearing techniques 
regarding the success metric.  Recall that a success occurs when a trace stops 
at a penultimate node.  The horizontal axis gives different filters size, from 
10,000 to 100,000, with an increment of 10,000. Below the horizontal axis sits 
another axis that indicates the compression ratio of the filter, compared to 
the list implementation of the RSS. The vertical axis gives the success rate.  
A value of 0 would mean that using a particular implementation precludes 
stopping at the penultimate node. On the other hand, a value of 1 means that 
the implementation succeeds in stopping each time at the penultimate node.

We see, from Fig.~\ref{case.compare.success}, that Improved Ratio Selection
performs better than standard Ratio Selection.  For $\beta=0.01$, the increase
is more important for larger vector size while it is the contrary for
$\beta=0.25$.

\begin{figure}[!t]
  \begin{center}
    \includegraphics[width=6.5cm]{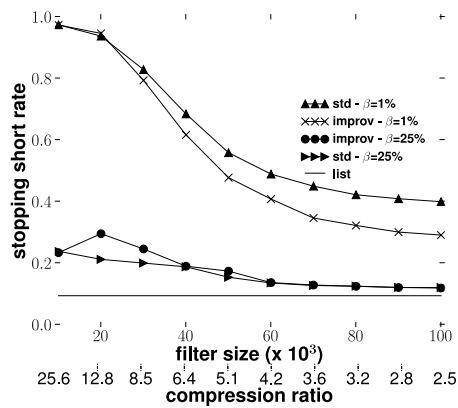}
  \end{center}
  \caption{Stopping short comparison}
  \label{case.compare.ss}
\end{figure}

Fig.~\ref{case.compare.ss} compares both selective clearing techniques 
regarding the stopping short metric.  Recall that a stopping short corresponds 
to an early stop in the probing, i.e., before the penultimate node. 

Again, we see from Fig.~\ref{case.compare.ss} that the improved selective
clearing algorithms performs better than standard algorithms.  This is more
explicit when $\beta=0.01$ and the vector is large.  However, for $\beta=0.25$,
we notice a small increase in the stopping short rate for some vector sizes
(between 20,000 and 30,000).

\begin{figure}[!t]
  \begin{center}
    \includegraphics[width=6.5cm]{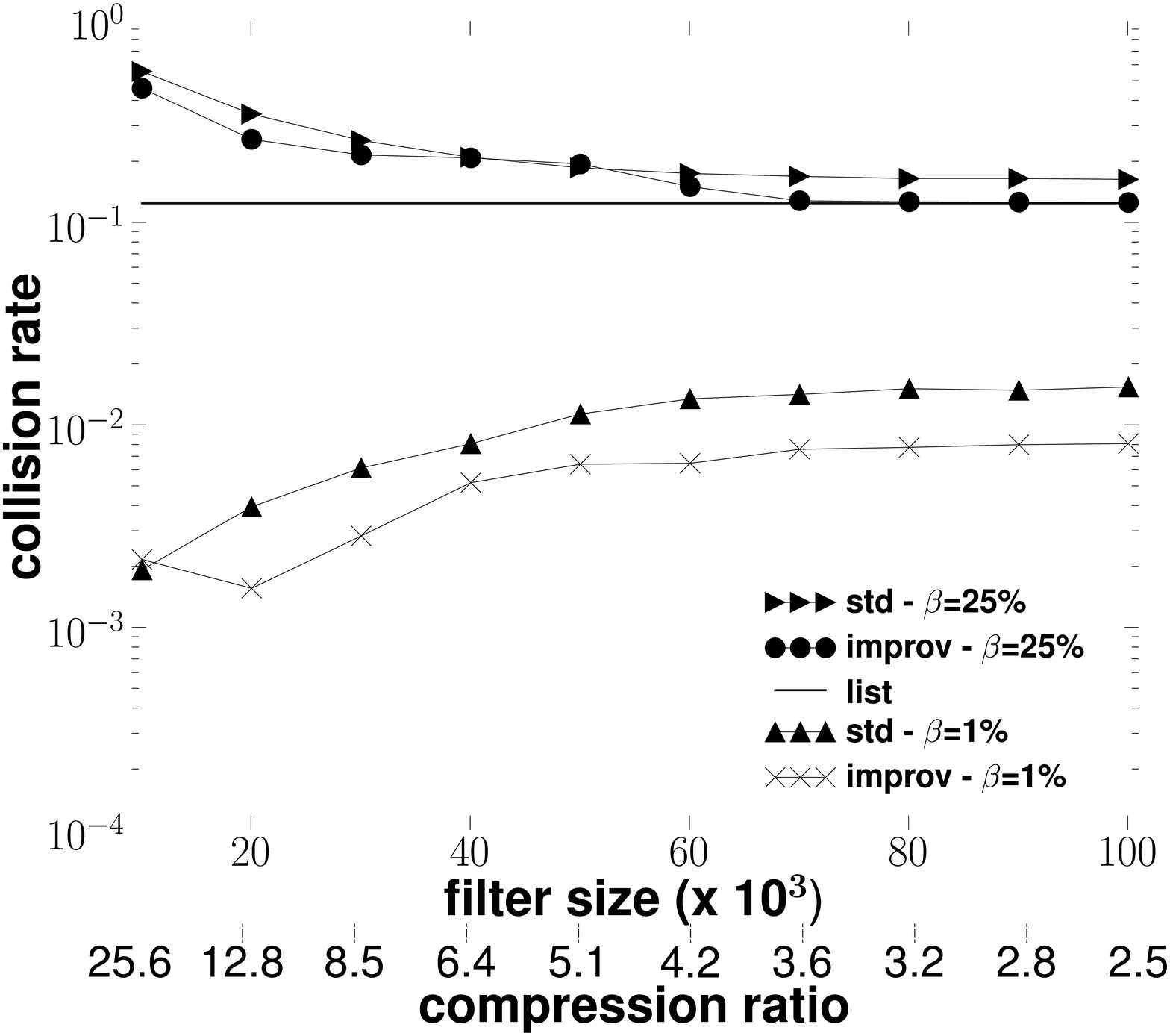}
  \end{center}
  \caption{Collision comparison}
  \label{case.compare.collision}
\end{figure}

Fig.~\ref{case.compare.collision} compares both selective clearing techniques
regarding the collision metric.  Recall that a collision occurs when a trace 
hits a destination. 

We note, from Fig.~\ref{case.compare.collision}, that Improved Ratio Selection
can decrease the collision rate compared to standard Ratio Selection.  We
further see that the collision rate, for Improved Ratio Selection, is very close
to the list one when the vector size is higher than 70,000 bits.  There is a
very tight difference between Improved Ratio Selection and the list that is not
visible on Fig.~\ref{case.compare.collision}.

In this section, we showed that using our improved selective clearing algorithms
can improve the performances of the RSS application.  Further, this performance
increase allows one to more reduce the size of the bit vector, leading to a
better compression ratio.

\section{Related Work}\label{related}
Early suggestions of applications for Bloom filters were for dictionaries and
databases. Bloom's original paper~\cite{bloom} describes their use for 
hyphenation.  Another dictionary application is for spell-checkers
\cite{spell1, spell2}. For databases, they have been suggested to speed up
semi-join operations \cite{db1, db2} and for differential files \cite{diff1, 
diff2}.

In this section, we discuss related work.  Our approach is double: first, we 
discuss Bloom filters variations and develop those that allow false negatives 
to arise (Sec.~\ref{related.bf}).  Second, we discuss networking applications
of Bloom filters and show that RBFs can find a suitable usage for some of them
(Sec.~\ref{related.app})

\subsection{Bloom Filters Variations}\label{related.bf}
\subsubsection{Extensions}\label{related.bf.extensions}
\dfn{Time-Decaying Bloom filters} (TBF), proposed by Cheng et 
al.~\cite{timeDecaying}, are somewhat similar to counting Bloom 
filters~\cite{counting} (CBF) as the standard bit vector is replaced by an 
array of counters.  TBF differs from CBF as values in the array decay 
periodically with time elapsing. TBF are used for maintaining time sensitive 
profiles of the web.  As only a small proportion of web content are frequently 
visited,  Cheng et al. propose that only heavy hitters are monitored by large 
counters, in order to avoid allocation larger counters to small values.

Chang et al.'s extension aims at supporting multiple binary predicates as
opposed to single binary predicate (the key $x$ belongs or not to $A$) of a
traditional Bloom filter~\cite{agingBF}.  Such an extension is needed in
packet classification, for instance, where a packet can be classified into
many, possibly disjoint, sets.  If the considered application required $I$ 
different sets, each cell of the bit vector will contained $I$ bits where the
$i^{th}$ bit in a cell corresponds to the $i^{th}$ set.  When the filter is
queried for a key membership, the $k$ bit strings returned by the hash 
functions are AND.  In the resulting bit string, if the $i^{th}$ bit is set to
1, it means that the key might belong to $i^{th}$ set.  The case where more
than one bit is set to 1 after the AND is not addressed by Chang et al.

The \dfn{time-out Bloom filters}~\cite{timeoutBF}, developed by Kong et al.~in 
the context of packet sampling, is an extension to standard Bloom filters where 
the bit vector is replaced by a bucket vector, each bucket containing a 
timestamp.  A bucket time-out $t_0$ is associated to the time-out Bloom filter. 
A time-out Bloom filter allows one to determine if an incoming packet belongs 
to an active flow or it is the first packet of a new flow.  When a packet with 
a timestamp $t$ arrives, it is compared with the $k$ timestamps, $v[h_1(t)], 
v[h_2(t)], \ldots, v[h_k(t)]$.  If at least one of the $k$ timestamps recorded 
in the filter follows $t - v[h_i(t)] > t_0$ (i.e.,  the bucket is time-out), 
the packet is sampled. Otherwise, it is discarded.  After the comparison, all 
the $k$ positions in the vector are updated with $t$ even if the packet is not 
sampled and all other buckets in the vector are set to 0. In a time-out Bloom 
filter, a bucket getting time-out is equivalent to a standard Bloom filter 
having a bit to 0, while a non time-out is the same as a bit to 1 in a Bloom 
filter.  Due to false positives, an time-out Bloom filter does not guarantee 
that all first packets can be sampled.

Kirsch and Mitzenmacher show that only two hash functions are needed to
effectively implement a Bloom filter without any loss in the false positive
probability~\cite{2hash}.  It also leads to less computation. The idea is to
use two hash functions $h_1(x)$ and $h_2(x)$ for simulation additional hash
functions of the form $g_i(x)~=~h_1(x)~+~ih_2(x)$.

\dfn{Space-code Bloom filters} by Kumar et al.~\cite{spaceCode} and
\dfn{spectral Bloom filters} by Cohen and Matias~\cite{spectral} are
approximate representation of a multiset, which allows for querying ``How many
occurrences of $x$ are there in set $M$?''.  A multiset is a set in which each
member has a multiplicity, i.e., a natural number indicating the occurrence of a
member in the set.

Based on the observation that, in many applications, some popular elements are
queried much frequently than the others, Bruck et al. propose the \dfn{weighted
Bloom filters} (WBF)~\cite{weightedBF}.  If the query frequency or the
membership likelihood is not uniform over all the keys in the universe, the
traditional configuration of the Bloom filter does not give the optimal
performance, as we demonstrated in Sec.~\ref{case.eval}.  In a WBF, each key $e
\in U$ is assigned $k_e$ hash functions, where $k_e$ depends on the query
frequency of $e$ and its likelihood of being a member of $A$.  Each non-member
element has a different false positive probability.  The average false positive
probability of a WBF is given by the weighted sum over the queries frequencies
of the elements in the universe.  A key is assigned more hash functions if its
query frequency is high and its chance of being a member is low.  When the query
frequencies and the membership likelihoods are the same for all keys in $U$, a
WBF behaves like a traditional Bloom filter.

The WBF differs from our RBF as it tries to build the Bloom filter in such a 
way that it reflects the key distribution.  However, it is not clear how a WBF 
can be used has a message shared between distributed entities as each key is, 
potentially, assigned a different number of hash functions.  There is an 
additional storage information associated to a WBF while an RBF modifies the 
traditional Bloom filter without adding any information.

Standard Bloom filters and most of their extensions are approaches to represent
a static set, i.e., the size of $A$ does not evolve with time.  However, for
many applications, for instance large-scale and distributed systems, it is
difficult to foresee the threshold size for the set $A$.  It is possible that
the size of $A$ will exceed its initial size, $n_0$, during the execution of
the application.  It is thus difficult, even impossible, to maintain the false
positive rate and the false positive probability will exceed its threshold.
Consequently, the Bloom filter can become unusable under such a scenario.

Two extensions of the standard Bloom filters have been proposed in order to
support dynamic sets.  The first one, \dfn{split Bloom filters}~\cite{splitBF},
uses a constant $s \times m$ bit matrix to represent a set, where $s$ is a
constant and must be pre-defined according to the estimation of the maximum
value of set size.  The second one, \dfn{dynamic Bloom
filters}~\cite{dynamicBF} (DBF) proposed by Guo et al., also makes use of a $s 
\times m$ bit matrix but each of the $s$ rows is a standard Bloom filter.  The
creation process of a DBF is iterative.  At the starting, the DBF is a $1 
\times m$ bit matrix, i.e., it is composed of a single standard Bloom filter.
It supposed that $n_r$ elements are recorded in the initial bit vector, where
$n_r \leq n$.  As the size of $A$ grows during the execution of the
application, several keys must be inserted in the DBF.  When inserting a key
into the DBF, one must first get an active Bloom filter in the matrix.  A Bloom
filter is active when the number of recorded keys, $n_r$, is strictly less than
the current cardinality of $A$, $n$.  If an active Bloom filter is found, the
key is inserted and $n_r$ is incremented by one.  On the other hand, if there
is no active Bloom filter, a new one is created (i.e., a new row is added to
the matrix) according to the current size of $A$ and the element is added in
this new Bloom filter and the $n_r$ value of this new Bloom filter is set to
one.  A given key is said to belong to the DBF if the $k$ positions are set to
one in one of the matrix rows.  Guo et al.~also extend standard Bloom filters
and DBF for supporting set consisted of multi-attribute keys.

\subsubsection{Bloom Filters and False Negatives}\label{related.bf.fn}
Has nobody thought of the RBF before?  There is a considerable literature on
Bloom filters, and their applications in networking, that we discuss in
Sec.~\ref{related.app}.  In a few instances, suggested variants on Bloom
filters do allow false negatives to arise.  However, these variants do not
preserve the size of the standard Bloom filter, as RBFs do.  Nor have the false
negatives been the subject of any analytic or simulation studies.  In
particular, the possibility of explicitly trading off false positives for false
negatives has not been studied prior to the current work, and efficient means
for performing such a trade-off have not been proposed.

First is the \dfn{anti-Bloom filter}, which was suggested in non-peer reviewed
work~\cite{antibloom1,antibloom2}. An anti-Bloom filter is composed of a
standard Bloom filter plus a separate smaller filter that can be used to
override selected positive results from the main filter. When queried, a
negative result is generated if either the main filter does not recognize a key
or the anti-filter does.  The anti-Bloom filter requires more space than the
standard filter, but the space efficiency has not been studied. Nor have
studies been made of the impact of the anti-filter on the false positive rate,
or on the false negatives that would be generated.

Second, Fan et al.'s CBF replaces each cell of a Bloom filter's bit vector with
a four-bit counter, so that instead of storing a simple 0 or a 1, the cell
stores a value between 0 and 15~\cite{counting}.  This additional space allows
CBFs to not only encode set membership information, as standard Bloom filters
do, but to also permit dynamic additions and deletions to that information.
One consequence of this new flexibility is that there is a chance of generating
false negatives. They can arise if counters overflow.  Fan et al.~suggest that
the counters be sized to keep the probability of false negatives to such a low
threshold that they are not a factor for the application (four bits being
adequate in their case).  The possibility of trading off false positives for
false negatives is not entertained.

Third, Bonani et al.'s $d$-left CBF is an improvement on the CBF.  As with the
CBF, it can produce false negatives. It can also produce another type of error
called ``don't know''.  Bonani et al.\ conduct experiments in which they
measure the rates for the different kinds of errors, but here too there is no
examination of the possibility of trading off false positives against false
negatives.  The $d$-left CBF is more space-efficient than the CBF.  But CBFs
themselves require a constant multiple more space than standard Bloom filters,
and the question does not arise of comparing the space efficiency of $d$-left
CBFs with that of standard Bloom filters, as they serve different functions.

Four, Song et al.~\cite{extendedBF} propose an extension to the CBF, called the
\dfn{extended Bloom filter} (EBF), in order to support exact address prefix
matching for routing.  An array is associated to the CBF. Each cell of this
array contains the list of keys that are recorded in the corresponding cell in
the CBF. Song et al.~propose several techniques to reduce the memory cost of
the EBFs.  The EBFs are designed to achieved higher lookup performance within
high-speed routers.

With the EBFs, the false positives are removed by adding information to the
CBFs.  With the RBFs, by contrast, no information is added to remove the false
positives.  The cost of these removals is expressed in terms of false negatives
generated for the RBFs and in terms of increased memory usage for the EBFs.

Five, Laufer et al.~\cite{generalized} an extension to the standard Bloom
filter called the \dfn{generalized Bloom filter} (GBF).  With the GBF, one
moves beyond the notion that elements must be encoded with 1s, and that 0s
represent the absence of information.  A GBF starts out as an arbitrary vector
of both 1s and 0s, and information is encoded by setting chosen bits to either
0 or 1. As a result, the GBF is a more general binary classifier than the
standard Bloom filter.  One consequence is that it can produce either false
positives or false negatives.  Laufer et al.\ provide a careful
analysis~\cite{gbfTechRep} of the trade-offs between false positives and false
negatives.

A GBF employs two sets of hash functions, $g_1, \ldots, g_{k_0}$ and $h_1, 
\ldots, h_{k_1}$ to set and reset bits. To add an element $x$ to the GBF, the
bits at positions $g_1(x), \ldots, g_{k_0}(x)$ are set to 0 and the bits at
positions $h_1(x), \ldots, h_{k_1}(x)$ are set to 1.  In the case of a
collision between two hash values $g_i(x)$ and $h_j(x)$, the bit is set to 0.
The membership of an element $y$ is verified by checking if all bits at
$g_1(y), \ldots, g_{k_0}(y)$ are set to 0 and all bits at $h_1(y), \ldots, 
h_{k_1}(y)$ are set to 1.  If at least one bit is inverted, $y$ does not belong
to the GBF with a high probability. A false negative arises when at least one
bit of $g_1(y), \ldots, g_{k_0}(y)$ is set to 1 or one bit of $h_1(y), \ldots, 
h_{k_1}(y)$ is set to 0 by another element inserted afterwards. The rates of
false positives and false negatives in a GBF can be traded off by varying the
numbers of hash functions, $k_0$ and $k_1$, as well as other parameters such as
the size of the filter.

RBFs differ from GBFs in that they allow the explicit removal of selected false
positives.  RBFs also do so in a way that allows the overall error rate,
expressed as a combination of false positives and false negatives, to be
lowered as compared to a standard Bloom filter of the same size.  We note that
the techniques used to remove false positives from standard Bloom filters could
be extended to remove false positives from GBFs.  For a false positive key,
$x$, either one would set one of the bits $g_1(x), \ldots, g_{k_0}(x)$ to 1 or
one of the bits $h_1(x), \ldots, h_{k_1}(x)$ to 0.

Finally, \dfn{Distance-sensitive Bloom filters}, introduced by Kirsch and
Mitzenmacher~\cite{distanceBF}, consider the notion of an ``approximate Bloom
filter'', that allows approximate instead of exact matches under a distance
metric.  In other words, a distance-sensitive Bloom filter tries to answer the
following question: ``Is $x$, where $x \in U$, close to an element belonging to
$A$?''.  In a distance-sensitive Bloom filter, classic hash functions are
replaced by distance-sensitive hash functions.  A Distance-sensitive Bloom
filter allow false positives and false negatives.

One might think that there would be general binary classifiers similar to RBFs
in the domain of machine learning. It is usual in artificial intelligence to
make use of classifiers, such as neural networks, Bayesian classifiers (naive
or not), or support vector machines (SVMs)~\cite{mitchell, modern}.  However,
these classifiers differ from RBFs in that they classify based on feature or
attribute vectors. RBFs classify elements purely on the basis of their unique
keys.

\subsection{Networking Bloom Filters Applications}\label{related.app}
Bloom filters have been widely used in networking applications, as stated by
Broder and Mitzenmacher~\cite{survey}.  Broder and Mitzenmacher consider four
types of networking applications: overlays and peer-to-peer networks, resource
processing, packet routing and measurement.  All these fours categories are
developed below.  For each of them, we discuss the use of RBFs instead of
traditional Bloom filters.

\subsubsection{Overlays and Peer-to-Peer}\label{related.app.p2p}
For a node in a peer-to-peer file sharing system, keeping a list of objects
stored at all other nodes might be costly in terms of memory, but keeping Bloom
filters for all other nodes might be an attractive alternative.  This was
proposed by Cuenca-Acuna et al.~for their \dfn{PlanetP} system~\cite{planetp}.

PlanetP meets the two criteria for the use of RBFs.  First, the application can
identify false positives.  A node, through is own experience with the inability
to locate certain files at the expected nodes, can determine that the keys
corresponding to those files yield false positives.  Second, false negatives
are tolerated because not every node that stores a given object need be
identified.  In a file sharing system, the same object is typically stored in
multiple locations, and so the failure of one node to recognize some of the
locations for some of the objects should not pose a great problem, provided the
rate of such errors remains within reasonable bounds. The communications
savings that come from eliminating some false positives might well outweigh the
costs of missing some locations.

Byers et al.~\cite{content} propose an application for distributing large files
to many peers in overlay networks.  They suggest that peers may want to solve
\dfn{approximate set reconciliation} problems. The idea is to allow a peer A to
send to a peer B objects that B does not have.  Encoding the sets of objects as
Bloom filters allows for data compression.  B will send A its Bloom filter.
Testing its own set, element by element, against this Bloom filter allows A to
know the set of objects B does not have, and send them to B. Because of false
positives, not all objects that B needs will be sent, but most will.

Approximate set reconciliation clearly meets the second criterion for using
RBFs.  A low rate of false negatives in the RBF furnished by peer B would
result in peer A sending a small number of elements that B already possesses.
It is easy to imagine that the system designers would be willing to pay this
communications overhead price in order to ensure that B gets more of the
elements that it is missing.

A question arises, however, for the first criterion.  How does peer B identify
the false positives in the Bloom filter that it sends out? For this, it would
need to know the keys for the objects that it is missing.  For some
applications, this would not be possible.  But we could easily imagine many
applications where the keys are known.  For instance, B might know the contents
of a music catalog, but not have many of the songs in that catalog.  It could
identify the false positives in its RBF by testing the keys in the catalog one
by one.

Rhea et al.~\cite{attenuated} describe a probabilistic algorithm for routing
peer-to-peer resource location queries.  Each node in the network keeps an
array of Bloom filters, called an \dfn{attenuated Bloom filter}, for each
adjacent edge in the overlay topology. In the array for each edge, there is a
Bloom filter for each distance $d$, up to a maximum value, so that the $d^{th}$
Bloom filter in the array keeps track of resources available via $d$ hops
through the overlay network along that edge.  If it is deemed probable that the
resource that is being searched for is present, the query is routed to the
nearest neighbor.  This scheme would require the addition of feedback to
identify false positives.  If false positives could be identified, they could
be removed.  This might be worthwhile, as false negatives do not invalidate the
system.  The array of Bloom filters could be replaced by an array of RBFs,
bringing about a decrease in the false positive rate at the cost of a
comparatively small increase in the false negative rate.

\subsubsection{Resource Routing}\label{related.app.routing}
Czerwinski et al.~\cite{resourceRouting} describe a resource discovery
architecture called \dfn{Ninja}, that makes use of Bloom filters.  In our
judgment, Ninja would not be tolerant of false negatives, and is thus not a
candidate for using RBFs.

However, another resource routing application could benefit.  Rhea et
al.~\cite{attenuated} describe a probabilistic algorithm for routing
peer-to-peer resource location queries.  Each node in the network keeps an
array of Bloom filters, called an \dfn{attenuated Bloom filter}, for each
adjacent edge in the overlay topology. In the array for each edge, there is a
Bloom filter for each distance $d$, up to a maximum value, so that the $d^{th}$
Bloom filter in the array keeps track of resources available via $d$ hops
through the overlay network along that edge.  If it is deemed probable that the
resource that is being searched for is present, the query is routed to the
nearest neighbor.  This scheme would require the addition of feedback to
identify false positives.  If false positives could be identified, they could
be removed.  This might be worthwhile, as false negatives do not invalidate the
system.  The array of Bloom filters could be replaced by an array of RBFs,
bringing about a decrease in the false positive rate at the cost of a
comparatively small increase in the false negative rate.

\subsubsection{Network Packet Processing}\label{related.app.packet}
Dharmapurikar et al.~\cite{parallelBF} propose the use of Bloom filters for
detecting predefined signatures in packet payloads. They propose an
architecture of $W$ \dfn{parallel Bloom filters}, each Bloom filter focusing on
strings of a specified length.  If a string is found to be a member of any
Bloom filter, it is then declared as a possible matching signature.  To avoid
the risk of false positives, each matching signature is tested in an
\dfn{analyzer} which determines if the signature is truly a member of the set
$A$ or not. In other words, the analyzer contains all elements of $A$.  Bloom
filters are only used to discard elements not belonging to $A$.

At least one of the two criteria for using the RBFs is met in the process
described by Dharmapurikar et al.  The analyzer offers the opportunity to
identify false positives.  The application should obtain a gain in terms of
processing time by removing from the filters those false positives.  The second
criterion is application-specific. If a small rate of false negatives may be
tolerated, then RBFs are suitable.

\subsubsection{Measurement}\label{related.app.measure}
Bloom filters are also used in topology discovery.  Some authors of this paper
propose \dfn{Doubletree}~\cite{DonnetEfficient2005}, an efficient and
cooperative algorithm that aims to reduce redundancy, i.e., duplication of
effort, in tracerouting systems by taking into account the tree-like structure
of routes in the internet. Reducing the redundancy implies coordination between
Doubletree monitors by sharing information about what was previously
discovered.  To summarize this information shared, Donnet et al.~propose to
implement it using Bloom filters~\cite{DonnetPam2005}. Though it is difficult
for Doubletree monitors to meet the first criterion by detecting false
positives, we proposed, in this paper (see Sec.~\ref{case}), a variant of the
problem for which RBFs are well adapted.

\section{Conclusion}\label{ccl}

\section*{Acknowledgements}
Mr.~Donnet's work was partially supported by a SATIN grant provided by the 
E-NEXT doctoral school, by an internship at \textsc{Caida}, and by the European 
Commission-funded OneLab project. Mark~Crovella introduced us to Bloom filters 
and encouraged our work.  Rafael P.~Laufer suggested useful references 
regarding Bloom filter variants.  Otto Carlos M.~B.~Duarte helped us clarify 
the relationship of RBFs to such variants.  We thank k~claffy and her team at 
\textsc{Caida} for allowing us to use the skitter data.

\bibliographystyle{IEEEtran}
\bibliography{rbf-TechRep}

\end{document}